\documentclass[11pt]{article}
\pdfoutput=1
\usepackage{amsmath} % for typesetting matrices
\usepackage{amssymb}
\usepackage{bm} % bold math
\usepackage{graphicx}
\usepackage[colorlinks=true,citecolor=blue,linkcolor=black]{hyperref}
\numberwithin{equation}{section}

\def\x{{\bf x}}
\def\k{{\bf k}}

\def\v{{\bf v}}

\def\vs{v_s}
\def\coeff#1#2{{\textstyle {\frac {#1}{#2}}}}
\def\u{\underline}

\begin{document}

\title{\Large First-order relativistic hydrodynamics is stable}
\author{\normalsize Pavel Kovtun\\
{\it\normalsize Department of Physics \& Astronomy,  University of Victoria,}\\
{\it\normalsize PO Box 1700 STN CSC, Victoria, BC,  V8W 2Y2, Canada}
}

\date{\normalsize July, 2019}

\maketitle

\begin{abstract}
\noindent
We study linearized stability in first-order relativistic viscous hydrodynamics in the most general frame. There is a region in the parameter space of transport coefficients where the perturbations of the equilibrium state are stable. This defines a class of stable frames, with the Landau-Lifshitz frame falling outside the class. The existence of stable frames suggests that viscous relativistic fluids may admit a sensible hydrodynamic description in terms of temperature, fluid velocity, and the chemical potential only, i.e.\ in terms of the same hydrodynamic variables as non-relativistic fluids. Alternatively, it suggests that the Israel-Stewart and similar constructions may be unnecessary for a sensible relativistic hydrodynamic theory.
\end{abstract}

\section{Introduction}
Do the Navier-Stokes equations of fluid dynamics admit a relativistic formulation? The answer to this question is subtle. The standard hydrodynamic equations of normal non-relativistic fluids represent local conservation laws of energy, momentum, and particle number.%
\footnote{
More generally, one needs conservation laws for the number of each species of particles. If only a single species of particles is present in a Galilean-invariant theory, the particle number conservation is equivalent to mass conservation. In what follows, we assume a single particle species in a Galilean-invariant theory, or a single global $U(1)$ charge in a Lorentz-invariant theory.}
The dynamical variables of hydrodynamics are the parameters that specify the thermal equilibrium state (temperature $T$, fluid velocity $\v$, and the chemical potential $\mu$), now promoted to functions of space and time.%
\footnote{
In the standard non-relativistic fluid dynamics, one may choose to use the energy density and the particle number density as the hydrodynamic variables, instead of $T$ and $\mu$. In a relativistic theory, energy density and the $U(1)$ charge density are not Lorentz scalars, so the covariant energy-momentum tensor and the $U(1)$ current are normally constructed in terms of $T$, $\mu$ (which are Lorentz scalars), and the fluid velocity $u^\alpha$ (which is a Lorentz vector).
}
The currents, including the dissipative fluxes, are expressed through the dynamical variables in terms of the phenomenological constitutive relations which contain the viscosities and the heat conductivity.

Let us ask the question in the following way: Do the hydrodynamic equations admit a relativistic formulation that (A) only uses the dynamical variables $T$, $u^\alpha$, and $\mu$ inherited from thermodynamics (similar to the non-relativistic Navier-Stokes fluids), and (B) gives rise to sensible physics, e.g.\ the equilibrium state is stable, and there is no superluminal propagation?

Phrased this way, the standard answer to this question was, until recently, a {\em no}. The original constructions of relativistic hydrodynamics by Eckart~\cite{PhysRev.58.919, weinberg:1972}, and by Landau and Lifshitz~\cite{LL6} predict that the thermal equilibrium state of a generic relativistic system is unstable~\cite{Hiscock:1985zz}, and moreover that there are modes that propagate faster than light~\cite{Hiscock:1987zz}. Thus the original hydrodynamic theories respect (A) but not (B). The unphysical features of the naive hydrodynamic equations are due to high-momentum, high-frequency modes. The problems with stability and causality may be rectified by introducing extra dynamical degrees of freedom into hydrodynamics which modify the behaviour of the theory at high momenta and high frequencies. This is what is done in the Israel-Stewart theory~\cite{Israel:1976tn, Israel-Stewart}, and in more modern derivations of the analogous stable and causal hydrodynamics such as \cite{Hiscock:1983zz, Baier:2007ix, Denicol:2012cn}. These theories thus respect (B) but not (A). Similarly, ``divergence-type'' formulations of hydrodynamics~\cite{Geroch:1990bw} respect (B) but not (A). Using an analogy with field theory, the extra dynamical degrees of freedom (besides $T$, $u^\alpha$ and $\mu$) play the role of an ultraviolet regulator, while keeping the low-energy physics in terms of $T$, $u^\alpha$ and $\mu$ intact. See~\cite{Rezzolla-Zanotti, Romatschke:2017ejr} for recent discussions of the formulation and applications of relativistic hydrodynamics. 

One may argue that preserving the condition (A) above is not crucial. After all, the Israel-Stewart and similar theories already provide viable ultraviolet regulators of the naive relativistic hydrodynamics. Nevertheless, not all ultraviolet regulators are the same: some may be more physical, and some may be easier to implement technically. At the very least, an ultraviolet completion of hydrodynamics that satisfies (A) has an aesthetic appeal similar to the dimensional regularization in perturbative field theory: a covariant regularization procedure without introducing extra degrees of freedom beyond those already present at low energies. Given the recent resurgence of interest in viscous relativistic hydrodynamics due to its successful application to flows of hot subnuclear matter in heavy-ion collisions~\cite{Romatschke:2017ejr, Gale:2013da, Jeon:2015dfa}, exploring different ultraviolet completions of the naive hydrodynamic theories deserves further attention.

The idea we would like to explore is whether both stability and causality might be maintained if one uses a certain out of equilibrium definition of the hydrodynamic variables which differs from the choice adopted by either Eckart or by Landau and Lifshitz. The improved behaviour of first-order relativistic hydrodynamics with non-Landau, non-Eckart definitions was touched upon in Refs.~\cite{Van:2011yn, Freistuhler20140055, Freistuhler20160729}, and at least for conformal fluids it appears that there {\em is} a formulation of first-order relativistic hydrodynamics that satisfies both (A) and (B)~\cite{Bemfica:2017wps}. Motivated by these developments, we will study the stability of first-order hydrodynamics with a general definition of the out of equilibrium variables. 

We will study viscous relativistic hydrodynamics in the most general frame.%
\footnote{
Following the standard terminology, by ``frame'' we mean a particular choice of how one defines $T$, $u^\alpha$ and $\mu$ out of equilibrium. Thus a change of ``frame'' is just a field redefinition of $T$, $u^\alpha$ and $\mu$ by derivative corrections. This use of the word ``frame'', while somewhat unfortunate, is widely adopted in the literature.
} 
Hydrodynamics is constructed as a derivative expansion, with gradients of $T$, $u^\alpha$ and $\mu$ parametrizing deviations from equilibrium. We will start with the most general constitutive relations to one-derivative order. Our emphasis here is not on the derivative expansion per se, but on whether the constitutive relations, truncated at the one-derivative order, can give rise to stable and causal hydrodynamic equations.

We don't use any particular model of matter. We also don't use the entropy current as a guiding principle.
One can reasonably demand that on-shell (i.e.\ when evaluated on the solutions to the equations of motion) the divergence of the entropy current $\nabla{\cdot}S$ must be non-negative order by order in the derivative expansion. This has been systematically implemented to second order in derivatives only relatively recently~\cite{Bhattacharyya:2012nq, Bhattacharyya:2013lha}. In first-order hydrodynamics one can only demand that on-shell $\nabla{\cdot}S\geqslant0$ to first order in derivatives: higher-order contributions to $\nabla{\cdot}S$ have to be resolved by higher-order hydrodynamics.
The general-frame first-order hydrodynamics that we study here has on-shell $\nabla{\cdot}S\geqslant0$ to first order in derivatives, as it should.%
\footnote{
In the kinetic theory derivation of the general-frame first-order hydrodynamics~\cite{Bemfica:2017wps} there always exists an entropy current with a non-negative divergence, by virtue of the Boltzmann's H-theorem.}
We review the hydrodynamic entropy current in the appendix.

The paper is organized as follows. We start with a detailed introduction to hydrodynamic frames in Sec.~\ref{sec:const-rel}, and explain how one arrives at the ``standard'' frames of Eckart and of Landau and Lifshitz. The most general frame in one-derivative hydrodynamics is specified by seven parameters for uncharged fluids (two of which are genuine one-derivative transport coefficients), and by sixteen parameters for charged fluids (three of which are genuine one-derivative transport coefficients). Section~\ref{sec:stability} contains a brief discussion of how we check for the stability of the global uniform equilibrium state. Then in Sec.~\ref{sec:uncharged} we restrict our attention to uncharged fluids and perform an analysis of small fluctuations in the equilibrium state with constant $T = T_0$ and $\v = \v_0$, and identify the class of hydrodynamic frames in which the fluctuations are stable. The Landau-Lifshitz frame (as well as the frames studied in~\cite{Hiscock:1985zz}) are outside this class. Entropy current is discussed in appendix~\ref{sec:entropy-current}, and the constraints on the general-frame transport coefficients in conformal fluids are discussed in appendix~\ref{sec:conformal}.

\section{Constitutive relations}
\label{sec:const-rel}
\subsection{Derivative expansion}
In order to set the stage for the further discussion, we start with a brief introduction to hydrodynamics along the lines of Ref.~\cite{Kovtun:2012rj}. Hydrodynamics is a classical effective theory for the evolution of the conserved densities, including energy density, momentum density, and various charge densities. We will be considering relativistic theories whose only relevant conserved densities are the energy-momentum tensor $T^{\mu\nu}$, and possibly a current $J^\mu$ corresponding to a global $U(1)$ symmetry. We will call the fluids without the corresponding global $U(1)$ current ``uncharged'', for example a $\varphi^4$ field theory of a real scalar field $\varphi$, or a ``pure glue'' $SU(N)$ Yang-Mills theory give rise to uncharged fluids. Similarly, the fluids with a conserved global $U(1)$ current will be called ``charged'', for example a $|\varphi|^4$ field theory of a complex scalar field $\varphi$, or quantum chromodynamics (QCD) give rise to charged fluids. For QCD, the relevant $U(1)$ charge is the baryon number.%
\footnote{
We emphasize that the term ``charged fluid'' used here, while standard in some discussions of relativistic hydrodynamics~\cite{Romatschke:2017ejr, Bhattacharyya:2012nq}, does not mean that the fluid has a net electric charge. Rather, we will use the term ``charged fluids'' to refer to fluids that can have non-zero local density of the baryon number, or other global $U(1)$ charges. The corresponding gauge field $A_\mu$ is external and non-dynamical. For fluids whose constituents carry actual electric charges (when the $U(1)$ is gauged, and $A_\mu$ is dynamical), the hydrodynamic description is provided by magneto-hydrodynamics~\cite{Hernandez:2017mch}. 
}

The expectation values $T^{\mu\nu}$, $J^\mu$ of the microscopic operators $\hat T^{\mu\nu}$, $\hat J^{\mu}$ are physical measurable quantities which can in principle be defined in any non-equilibrium state. In equilibrium, the state of the system in the grand canonical ensemble may be parametrized by the temperature $T$, the timelike velocity vector $u^\alpha$, and by the chemical potential $\mu$ for the $U(1)$ charge. The equilibrium energy-momentum tensor and the current can be expressed in terms of $T$, $u^\alpha$, and $\mu$. For states not too far out of equilibrium, the hydrodynamic assumption asserts that the physical objects $T^{\mu\nu}=\langle \hat T^{\mu\nu}\rangle$, $J^{\mu}=\langle \hat J^{\mu}\rangle$ can still be expressed in terms of the quantities $T$, $u^\alpha$ and $\mu$ that vary slowly in space and time. In equilibrium, the quantities $T$, $u^\alpha$ and $\mu$ become the actual temperature, fluid velocity, and the chemical potential. However, out of equilibrium, $T$, $u^\alpha$ and $\mu$ have no first-principles microscopic definitions, and thus should be viewed as merely auxiliary variables used to parametrize the physical observables $T^{\mu\nu}$ and $J^\mu$. 

Assuming that the physical system is locally near thermal equilibrium, the hydrodynamic expansion is a gradient expansion, schematically%
\begin{subequations}
\label{eq:TJ-expansion-1}
\begin{align}
  & T^{\mu\nu} = {\cal O}(1) + {\cal O}(\partial) + {\cal O}(\partial^2) + {\cal O}(\partial^3) + \dots\,,\\
  & J^{\mu} = {\cal O}(1) + {\cal O}(\partial) + {\cal O}(\partial^2) + {\cal O}(\partial^3) + \dots\,,
\end{align}
\end{subequations}
where ${\cal O}(\partial^k)$ denotes the terms with $k$ derivatives of $T$, $u^\alpha$, $\mu$, for example the ${\cal O}(\partial^2)$ contributions contain terms proportional to $\partial^2 T$, $(\partial T)^2$, $(\partial T)(\partial u)$ etc. Expansions (\ref{eq:TJ-expansion-1}) are the constitutive relations, expressing the physical quantities $T^{\mu\nu}$ and $J^\mu$ in terms of the auxiliary variables $T$, $u^\alpha$, $\mu$. The ${\cal O}(1)$ terms in the expansion are usually said to correspond to ``perfect fluids'', the ${\cal O}(\partial)$ terms are said to correspond to ``viscous fluids'', and the ${\cal O}(\partial^k)$ terms are said to correspond to ``$k^{\rm th}$-order hydrodynamics''. The hydrodynamic equations are $\partial_\mu T^{\mu\nu}=0$, $\partial_\mu J^\mu = 0$. In what follows, we will assume that the fluid is in flat space in 3+1 dimensions, and is not subject to external gauge fields that couple to the $U(1)$ current. The constitutive relations in curved space and in the presence of external gauge fields are known up to ${\cal O}(\partial^2)$~\cite{Bhattacharyya:2012nq, Bhattacharyya:2014bha}, but we will not need them here.

The derivative expansion can be implemented in practice in the following way. Given a timelike unit vector $u^\mu$, the energy-momentum tensor and the current may be decomposed as%
\begin{subequations}
\label{eq:T1}
\begin{align}
  &  T^{\mu\nu} = {\cal E} u^\mu u^\nu + {\cal P} \Delta^{\mu\nu} +
     ({\cal Q}^\mu u^\nu + {\cal Q}^\nu u^\mu) + {\cal T}^{\mu\nu}\,,\\
  & J^\mu = {\cal N} u^\mu + {\cal J}^\mu\,,
\end{align}
\end{subequations}
where ${\cal Q}^\mu$, ${\cal T}^{\mu\nu}$, and ${\cal J}^\mu$ are transverse to $u$, and ${\cal T}^{\mu\nu}$ is symmetric and traceless. Specifically,%
\begin{subequations}
\label{eq:EPNqjt}
\begin{align}
  & {\cal E} \equiv u_\mu u_\nu T^{\mu\nu}\,,\ \ \ \ 
     {\cal P} \equiv \frac{1}{d} \Delta_{\mu\nu} T^{\mu\nu}\,,\ \ \ \ 
     {\cal Q}_\mu \equiv -\Delta_{\mu\alpha} u_\beta T^{\alpha\beta}\,,\\[5pt]
  & {\cal T}_{\mu\nu} \equiv \frac12\left(
     \Delta_{\mu\alpha}\Delta_{\nu\beta} + \Delta_{\nu\alpha}\Delta_{\mu\beta}
     -\frac{2}{d} \Delta_{\mu\nu} \Delta_{\alpha\beta}
     \right) T^{\alpha\beta}\,,\\[5pt]
  & {\cal N} \equiv - u_\mu J^\mu\,,\ \ \ \ 
    {\cal J}_\mu \equiv \Delta_{\mu\alpha} J^\alpha\,,
\end{align}
\end{subequations}
where $\Delta^{\alpha\beta}\equiv g^{\alpha\beta}+u^\alpha u^\beta$ projects onto the space orthogonal to $u_\alpha$. The decompositions~(\ref{eq:T1}) are just identities, true for any symmetric tensor $T^{\mu\nu}$ and any vector $J^\mu$. In hydrodynamics, one writes ${\cal E}$, ${\cal P}$, ${\cal Q}^\mu$, ${\cal T}^{\mu\nu}$, ${\cal N}$, and ${\cal J}^\mu$ as a derivative expansion. To zeroth order in derivatives, there are two scalars, $T$ and $\mu$, no transverse vectors, and no transverse traceless 2-tensors. To first order in derivatives, there are three scalars, $\dot T$, $\partial_\lambda u^\lambda$, and $\dot\mu$, where the dot stands for $u^\lambda \partial_\lambda$. There are also three transverse vectors, $\Delta^{\rho\sigma} \partial_\sigma T$, $\dot u^\rho$, and $\Delta^{\rho\sigma} \partial_\sigma \mu$. There is one transverse traceless symmetric tensor, $\sigma^{\mu\nu} = \Delta^{\mu\rho}\Delta^{\nu\sigma}(\partial_\rho u_\sigma + \partial_\sigma u_\rho - \frac23 g_{\rho\sigma} \partial_\lambda u^\lambda)$. Thus to first order in derivatives we have
\begin{subequations}
\label{eq:EPQTNJ}
\begin{align}
 {\cal E} & = \epsilon + \varepsilon_1 \dot T/T + \varepsilon_2 \partial_\lambda u^\lambda + \varepsilon_3 u^\lambda \partial_\lambda(\mu/T) + O(\partial^2)\,, \\[5pt]
 {\cal P} & = p + \pi_1 \dot T/T + \pi_2 \partial_\lambda u^\lambda +\pi_3 u^\lambda \partial_\lambda(\mu/T) + O(\partial^2)\,, \\[5pt]
 {\cal Q}^\mu & = \theta_1  \dot u^\mu + \theta_2/T \,\Delta^{\mu\lambda} \partial_\lambda T + \theta_3 \Delta^{\mu\lambda} \partial_\lambda (\mu/T) +  O(\partial^2)\,,\\[5pt]
 {\cal T}^{\mu\nu} & = -\eta \sigma^{\mu\nu} + O(\partial^2)\,,\\[5pt]
 {\cal N} & = n + \nu_1 \dot T/T + \nu_2 \partial_\lambda u^\lambda + \nu_3 u^\lambda \partial_\lambda(\mu/T) + O(\partial^2)\,, \\[5pt]
 {\cal J}^\mu & = \gamma_1  \dot u^\mu + \gamma_2/T \,\Delta^{\mu\lambda} \partial_\lambda T + \gamma_3 \Delta^{\mu\lambda} \partial_\lambda (\mu/T) +  O(\partial^2)\,,
\end{align}
\end{subequations}
where $O(\partial^k)$ denotes terms with $k$ or more derivatives of the hydrodynamic variables. The factors of $1/T$ are inserted for notational convenience.
At zero-derivative order, the constitutive relations are determined by the three apriori independent parameters $\epsilon$, $p$, and $n$ which in general all depend on $T$ and $\mu$. 
As usual, $p$ is the pressure, $\epsilon$ is the energy density, and $n$ is the charge density.
At one-derivative order, there are sixteen apriori independent transport coefficients (seven for uncharged fluids) $\varepsilon_{1,2,3}$, $\pi_{1,2,3}$, $\theta_{1,2,3}$, $\nu_{1,2,3}$, $\gamma_{1,2,3}$, and $\eta$, which in general all depend on $T$ and $\mu$. Not all of them are genuine one-derivative transport coefficients though. As we will see shortly, there are in fact only three genuine one-derivative transport coefficients.

\subsection{Field redefinitions}
Out of equilibrium, the auxiliary variables $T$, $u^\alpha$ and $\mu$ have no first-principles microscopic definition, and different out-of-equilibrium choices of $T$, $u^\alpha$ and $\mu$ may be adopted, as long as all these choices agree in equilibrium. Given a choice of $T$, $u^\alpha$ and $\mu$, we can introduce 
\begin{equation}
\label{eq:frame-change}
  T' = T+\delta T\,,\ \ \ \ u'^\alpha = u^\alpha + \delta u^\alpha\,,\ \ \ \ \mu' = \mu + \delta\mu\,,
\end{equation}
where $\delta T$, $\delta u^\alpha$, $\delta\mu$ are $O(\partial)$. In this way, $T$ and $T'$, $u^\alpha$ and $u'^\alpha$, $\mu$ and $\mu'$ agree in equilibrium in flat space.%
\footnote{
This assumes that a certain equilibrium definition of what one means by $T$, $u^\alpha$, and $\mu$ has been adopted. In curved space, different definitions of {\em equilibrium} $T$, $u^\alpha$ and $\mu$ exist in the literature. In principle, within the derivative expansion, it is legitimate to consider field redefinitions that include non-equilibrium as well as equilibrium terms, such as contributions to $\delta T$, $\delta u^\alpha$ and $\delta\mu$ that depend on spatial derivatives of the external sources (metric and the gauge field). 
}
Note that $u_\mu \delta u^\mu = O(\partial^2)$ in order to maintain the normalization $(u')^2=-1$. The same $T^{\mu\nu}$ and $J^\mu$ may now be written in terms of the new (primed) variables.%
\footnote{
The freedom to redefine the hydrodynamic variables in the derivative expansion implies that the notions of ``local temperature'', ``local chemical potential'', and ``local rest frame of the fluid'' are intrinsically ambiguous. What is not ambiguous are the energy-momentum tensor and the current.
}
Again, when hydrodynamics is formulated as a derivative expansion, we have
\begin{subequations}
\begin{align}
  & T^{\mu\nu} = {\cal O}(1) + {\cal O}(\partial T',\partial u',\partial\mu') + O(\partial^2)\,,\\[5pt]
  & J^\mu = {\cal O}(1) + {\cal O}(\partial T',\partial u',\partial\mu') + O(\partial^2)\,.
\end{align}
\end{subequations}
The energy-momentum tensor and the current can be decomposed as in Eq.~(\ref{eq:T1}), (\ref{eq:EPNqjt}), now with respect to the new fluid velocity $u'^\mu$. The coefficients in the new decomposition are~\cite{Kovtun:2012rj}%
\begin{subequations}
\begin{align}
  & {\cal E}'(T',\mu') = {\cal E}(T,\mu) + O(\partial^2)\,,\\ 
  & {\cal P}'(T',\mu') = {\cal P}(T,\mu) + O(\partial^2)\,,\\
  & {\cal Q}'_\mu(T',u',\mu') = {\cal Q}_\mu(T,u,\mu) - (\epsilon{+}p)\delta u_\mu + O(\partial^2)\,,\\
  & {\cal T}'_{\mu\nu}(T',u',\mu') = {\cal T}_{\mu\nu}(T,u,\mu) + O(\partial^2)\,,\\
  & {\cal N}'(T',\mu') = {\cal N}(T,\mu) + O(\partial^2)\,,\\
  & {\cal J}'_\alpha(T',u',\mu') = {\cal J}_\alpha(T,u,\mu) - n\, \delta u_\alpha + O(\partial^2)\,.
\end{align}
\end{subequations}
Suppose now that we perform the most general first-order field redefinition
\begin{subequations}
\label{eq:frame-change-2}
\begin{align}
  & \delta T = a_1 \dot T/T + a_2 \partial{\cdot}u+ a_3 u^\lambda \partial_\lambda(\mu/T)\,,\\
  & \delta u^\mu =  b_1 \dot u^\mu + b_2/T\, \Delta^{\mu\nu}\partial_\nu T + b_3 \Delta^{\mu\lambda} \partial_\lambda(\mu/T)\,,\\
  & \delta \mu = c_1 \dot T/T + c_2 \partial{\cdot}u+ c_3 u^\lambda \partial_\lambda(\mu/T)\,,
\end{align}
\end{subequations}
with yet unspecified $a_i$, $b_i$, $c_i$ which are functions of $T$ and $\mu$. The constitutive relations for the energy-momentum tensor and the current, written in terms of the new fields $T'$, $u'$, $\mu'$, look the same as the constitutive relations in terms of the old fields $T$, $u$, $\mu$, with the following change:%
\begin{subequations}
\label{eq:frame-change-3}
\begin{align}
  & \varepsilon_i \to \varepsilon_i - \epsilon_{,T}a_i - \epsilon_{,\mu} c_i\,,\\
  & \pi_i \to \pi_i - p_{,T}a_i - p_{,\mu} c_i\,,\\
  & \nu_i \to \nu_i - n_{,T}a_i - n_{,\mu} c_i\,,\\
  & \theta_i \to \theta_i - (\epsilon{+}p)b_i\,,\\
  & \gamma_i \to \gamma_i - n b_i\,,\\
  & \eta \to \eta\,,
\end{align}
\end{subequations}
where $i=1,2,3$, and the comma subscript denotes the partial derivative with respect to the parameter that follows, e.g.\ $\epsilon_{,T} \equiv (\partial\epsilon/\partial T)_\mu$. Cleary, not all transport coefficients are invariant under the general first-order field redefinition. The invariant ones are 
\begin{align}
\label{eq:frame-invariants-1}
  f_i\equiv \pi_i - \left(\frac{\partial p}{\partial \epsilon}\right)_{\!n}\varepsilon_i - \left(\frac{\partial p}{\partial n}\right)_{\!\epsilon}\nu_i \,,\ \ \ \ \ \ 
  \ell_i \equiv \gamma_i - \frac{n}{\epsilon+p} \theta_i\,,
\end{align}
and $\eta$, while the rest can be eliminated from the first-order constitutive relations by a field redefinition. We see that when the constitutive relations (\ref{eq:EPQTNJ}) are truncated at one-derivative order by neglecting all $O(\partial^2)$ terms and taken at face value, there is not a unique first-order hydrodynamics. This non-uniqueness is due to the freedom of the hydrodynamic field redefinition (\ref{eq:frame-change-2}) which is conventionally called ``a change of frame''. 

The arbitrariness in the choice of frame is not important from the point of view of the derivative expansion. However, it can be important from the point of view of the hydrodynamic equations themselves. After all, the hydrodynamic equations $\partial_\mu T^{\mu\nu}=0$, $\partial_\mu J^\mu=0$ (again, with the constitutive relations truncated at one-derivative order), when written in different frames, give rise to {\em different} partial differential equations. The choice of frame may potentially affect such things as the well-posedness of the initial value problem for these partial differential equations, or lead to fictitious instabilities of the equilibrium state. 

Before discussing the conventional frames such as the Landau-Lifshitz frame or the Eckart frame, it is important to emphasize that no matter what one does in the hydrodynamics which includes $T$, $u^\alpha$, $\mu$ as dynamical variables, one always uses {\em some} frame. This is because a frame is just a definition of what one means by $T$, $u^\alpha$ and $\mu$ out of equilibrium, so ``not choosing a frame'' is not an option. The moment the constitutive relations for $T^{\mu\nu}$ and $J^{\mu}$ are written down in terms of $T$, $u^\alpha$ and $\mu$, with specified values for the $T$, $\mu$-dependent transport coefficients~--- these constitutive relations define a frame.

\subsubsection{Consequences of extensivity}
\label{sec:extensivity}
In equilibrium, extensivity follows from locality on length scales longer than the equilibrium correlation length. Extensivity implies certain thermodynamic consistency conditions that were introduced in Refs.~\cite{Banerjee:2012iz, Jensen:2012jh} in the context of the derivative expansion in hydrostatics when the relativistic system is subject to external time-independent sources (metric and the gauge field). The fluid velocity in the time-independent equilibrium state was chosen to be aligned with the corresponding timelike Killing vector. If the metric does not depend on time $t$ so that the Killing vector is $K=\partial_t$, the equilibrium temperature and the chemical potential are proportional to $1/\sqrt{-g_{00}}$ to all orders in the derivative expansion, in agreement with Ref.~\cite{LL5}. This might seem like the most natural thing to do, however it has non-trivial consequences. In particular, the Killing equation implies that the following identities hold in equilibrium:
\begin{subequations}
\begin{align}
  & \dot T = \dot\mu = 0\,,\\
  & T \dot u^\mu + \Delta^{\mu\nu}\partial_\nu T = 0\,,\\
  & E^\alpha - T \Delta^{\alpha\nu}\partial_\nu(\mu/T) = 0\,,\\
  & \nabla_{\!\mu}u^\mu = 0\,,\ \ \ \ \sigma^{\alpha\beta} = 0\,,
\end{align}
\end{subequations}
where $E^\alpha$ is the electric field, which in our case is set to zero. Note that in an external gravitational field, $\dot u^\mu$ and $\Delta^{\mu\nu}\partial_\nu T$ do not have to vanish separately in equilibrium. The thermodynamic consistency conditions arise from demanding that the equilibrium energy-momentum tensor and the current follow from the grand canonical potential which is extensive in the thermodynamic limit.
Varying the grand canonical potential (the generating functional) with respect to the metric and the gauge field one finds 
\begin{subequations}
\label{eq:TJ-thermo}
\begin{align}
  & \left. T^{\mu\nu}\right|_{\rm equilibrium} = \left(-p + T\frac{\partial p}{\partial T} + \mu\frac{\partial p}{\partial\mu} \right)  u^\mu u^\nu + p\Delta^{\mu\nu} + O(\partial^2)\,,\\
  & \left. J^\alpha \right|_{\rm equilibrium} = \left(\frac{\partial p}{\partial\mu}\right) u^\alpha + O(\partial^2)\,.
\end{align}
\end{subequations}
This implies that the coefficients in the constitutive relations (\ref{eq:EPQTNJ}) are not all independent. At zero-derivative order, the thermodynamic consistency conditions demand that $\epsilon$, $p$, and $n$ are not independent, but rather must satisfy $\epsilon = -p+T\frac{\partial p}{\partial T} +\mu \frac{\partial p}{\partial\mu}$, and $n=\partial p/\partial\mu$. These relations are of course well known from basic thermodynamics; what was not appreciated until \cite{Banerjee:2012iz, Jensen:2012jh} is that there are also relations analogous to $\epsilon+p=Ts+\mu n$ at ${\cal O}(\partial)$ and higher in the derivative expansion. At one-derivative order, the consistency of (\ref{eq:TJ-thermo}) and (\ref{eq:EPQTNJ}) implies $\theta_1 = \theta_2$, $\gamma_1 = \gamma_2$, and therefore $\ell_1 = \ell_2$. Analogous consistency conditions for two-derivative terms in the constitutive relations were worked out in~\cite{Banerjee:2012iz, Jensen:2012jh}.

Choosing the equilibrium fluid velocity along the Killing vector, $u = K/\sqrt{-K^2}$, and choosing $T = T_0/\sqrt{-K^2}$ (here $T_0$ is a constant that sets the units of temperature), the energy-momentum tensor and the current derived from the grand canonical potential were termed in Ref.~\cite{Jensen:2012jh} to be in ``thermodynamic frame''. We see however that this definition of $u$ and $T$ implies relationships among frame invariants, and therefore the thermodynamic frame is more than a mere choice of frame: it is a frame that makes the thermodynamic consistency conditions manifest.

\subsubsection{Landau-Lifshitz frame}
The Landau-Lifshitz frame~\cite{LL6} (also called the Landau frame) takes a different starting point. Consider the most general constitutive relations (\ref{eq:EPQTNJ}), with all transport coefficients non-zero. Given the transformation rules (\ref{eq:frame-change-3}), it is clear that one can go to a new frame (by choosing $a_i$ and $c_i$ appropriately) in which ${\cal E} = \epsilon$ and ${\cal N} = n$. Further, by choosing $b_i = \theta_i/(\epsilon{+}p)$ one can make ${\cal Q}^\mu = 0$ in the new frame. The constitutive relations in the new frame are then%
\begin{subequations}
\label{eq:Landau-frame-1}
\begin{align}
  & T^{\mu\nu} = \epsilon u^\mu u^\nu + \left[ p +f_1 \dot T/T + f_2 \partial{\cdot}u + f_3 u^\lambda \partial_\lambda(\mu/T) \right] \Delta^{\mu\nu} - \eta \sigma^{\mu\nu} + O(\partial^2)\,,\\
  & J^\mu = n u^\mu + \ell_1  \dot u^\mu + \ell_2/T \,\Delta^{\mu\lambda} \partial_\lambda T + \ell_3 \Delta^{\mu\lambda} \partial_\lambda (\mu/T) + O(\partial^2)\,,
\end{align}
\end{subequations}
where $f_i$ and $\ell_i$ are defined by Eq.~(\ref{eq:frame-invariants-1}). In other words, by performing a particular nine-parameter frame transformation the number of one-derivative transport coefficients has been reduced from sixteen to seven, which are now $f_{1,2,3}$, $\ell_{1,2,3}$, and $\eta$. The conditions ${\cal E} = \epsilon$, ${\cal N} = n$, and ${\cal Q}^\mu=0$ are usually taken as the defining properties of the Landau-Lifshitz frame.

Up to now we have only looked at the constitutive relations, but have not actually used the hydrodynamic equations themselves, $\partial_\mu T^{\mu\nu} = 0$, $\partial_\mu J^\mu = 0$. The conservation equations $\partial_\mu (n u^\mu) + O(\partial^2) = 0$ and $u_\nu \partial_\mu (\epsilon u^\mu u^\nu + p \Delta^{\mu\nu}) +O(\partial^2) = 0$ imply two ``on-shell'' relations among the scalars $\dot T$, $\partial{\cdot}u$, and $\dot\mu$, up to $O(\partial^2)$ terms. Similarly, the projected energy-momentum conservation $\Delta^\alpha_\nu \partial_\mu (\epsilon u^\mu u^\nu + p \Delta^{\mu\nu}) +O(\partial^2) = 0$ implies one ``on-shell'' relation among the transverse vectors $\dot u^\alpha$, $\Delta^{\alpha\lambda} \partial_\lambda T$, and $\Delta^{\alpha\lambda} \partial_\lambda (\mu/T)$, up to $O(\partial^2)$ terms. If these are used to eliminate $\dot T$, $\dot\mu$, and $\dot u^\mu$, the constitutive relations (\ref{eq:Landau-frame-1}), {\em when evaluated on the solutions to the hydrodynamic equations}, may be written as
\begin{subequations}
\label{eq:Landau-frame-2}
\begin{align}
  & T^{\mu\nu} = \epsilon u^\mu u^\nu + \left[ p -\zeta \partial{\cdot}u \right] \Delta^{\mu\nu} - \eta \sigma^{\mu\nu} + O(\partial^2)\,,\\
  & J^\alpha = n u^\alpha - \sigma T \Delta^{\alpha\nu}\partial_\nu (\mu/T) + \chi_{\rm T} \Delta^{\alpha\nu}\partial_\nu T + O(\partial^2)\,.
\end{align}
\end{subequations}
The transport coefficients include the shear viscosity $\eta$ as well as:
\begin{subequations}
\label{eq:zetasigma}
\begin{align} 
  & \zeta = -f_2 + \frac{\left[ (\epsilon{+}p)\frac{\partial n}{\partial\mu} -n \frac{\partial\epsilon}{\partial\mu} \right] f_1 + \left[ n\left(\frac{\partial\epsilon}{\partial T}\right)_{\mu/T} -  (\epsilon{+}p) \left(\frac{\partial n}{\partial T}\right)_{\mu/T}\right] f_3}{T(\frac{\partial\epsilon}{\partial T}\frac{\partial n}{\partial\mu} - \frac{\partial\epsilon}{\partial\mu} \frac{\partial n}{\partial T})} \,,
    \\[5pt]
  & \sigma = \frac{n}{\epsilon+p}\ell_1 - \frac{1}{T}\ell_3\,,\\[5pt]
  & \chi_{\rm T} = \frac{1}{T}\left( \ell_2 - \ell_1 \right)\,,
\end{align}
\end{subequations}
where the thermodynamic derivatives are $(\partial n/\partial T)_{\mu/T} = (\partial n/\partial T) + (\mu/T)(\partial n/\partial\mu)$, and the coefficients $f_i$, $\ell_i$ are given by Eq.~(\ref{eq:frame-invariants-1}). The above expressions give the bulk viscosity $\zeta$ and the charge conductivity $\sigma$ as linear combinations of the transport coefficients in the general frame.%
\footnote{
\label{fn:zeta}
For an uncharged fluid, the bulk viscosity is $\zeta = \vs^2(\pi_1 {-} \vs^2 \varepsilon_1) -\pi_2 + \vs^2\varepsilon_2$, where $\vs^2 \equiv \partial p/\partial\epsilon$ is the speed of sound squared in an uncharged fluid.
}
The transport coefficient $\chi_{\rm T}$ vanishes as a consequence of the thermodynamic consistency conditions $\ell_1 = \ell_2$. 
Thus one is left with three genuine one-derivative transport coefficients $\eta$, $\zeta$, and $\sigma$. The constitutive relations in the form (\ref{eq:Landau-frame-2}) with $\chi_{\rm T}=0$ were proposed in the book by Landau and Lifshitz~\cite{LL6}.

The above discussion makes it clear that the Landau-Lifshitz frame is not the only possible frame, {\em even for an uncharged fluid in one-derivative hydrodynamics}.%
\footnote{
The advantage of the Landau-Lifshitz frame is that it allows for a technically straightforward algebraic extraction of the hydrodynamic fields $T$ and $u^\mu$ if the energy-momentum tensor happens to be known. Indeed, suppose we know $T^{\mu\nu}(x)$ in some non-equilibrium state. For states that represent small departures from local equilibrium, we expect that $T^{\mu\nu}(x)$ will have a time-like eigenvector, i.e.\ for every $x$ we have $T^{\mu\nu}(x) \Psi_\nu(x) = \lambda(x)\Psi^\mu(x)$, where $\lambda(x)$ is the eigenvalue. After finding the eigenvector, the fluid velocity in the Landau-Lifshitz frame is just $u^\mu(x) = \Psi^\mu(x)/\sqrt{-\Psi^2}$. After finding the eigenvalue, the temperature $T(x)$ in the Landau-Lifshitz frame is found from $\lambda(x) = \epsilon(T(x))$, where $\epsilon(T)$ is a known function, given by the equilibrium equation of state. This procedure is algebraic, and can in principle be done independently at each point in spacetime. If, on the other hand, we want to match the exact known $T^{\mu\nu}(x)$ to the hydrodynamic form (\ref{eq:T1}), (\ref{eq:EPQTNJ}), with some fixed non-zero $\varepsilon_1$, $\varepsilon_2$, $\theta_1$, $\theta_2$, doing so would involve solving differential equations in order to extract $T(x)$ and $u^\mu(x)$, and so is technically more involved.
}
It is also important to note that the ingredients that go into the constitutive relations (\ref{eq:Landau-frame-2}) are: {\it (i)} the choice of frame such that ${\cal E} = \epsilon$, ${\cal N}=n$, ${\cal Q}^\mu=0$, plus {\it (ii)} the on-shell relations derived from zero-derivative hydrodynamics. The arbitrariness involved in the second step implies that there are multiple ways to write down the on-shell constitutive relations in the Landau frame -- these constitutive relations, when truncated to one-derivative order, will give rise to inequivalent hydrodynamic equations. Put differently, what is commonly referred to as the Landau frame is a whole class of frames. For example, if we use the on-shell relations to eliminate $\partial{\cdot}u$ and $\dot\mu$, the non-equilibrium correction to pressure in the Landau frame will be determined by $\dot T$ instead of $\partial{\cdot}u$, with the coefficient of $\dot T$ proportional to the bulk viscosity. Eliminating different one-derivative quantities will lead to different Landau-frame hydrodynamic equations which may have different stability and causality properties compared to the ones derived from~(\ref{eq:Landau-frame-2}).

\subsubsection{Eckart frame}
In order to arrive at the Eckart frame~\cite{PhysRev.58.919}, again consider the general constitutive relations (\ref{eq:EPQTNJ}), with all transport coefficients non-zero. Using the transformation rules (\ref{eq:frame-change-3}),  one can go to a new frame (by choosing $a_i$ and $c_i$ appropriately) in which ${\cal E} = \epsilon$ and ${\cal N} = n$, just like in the Landau frame. Further, by choosing $b_i = \gamma_i/n$ one can make ${\cal J}^\mu = 0$ in the new frame. The constitutive relations in the new frame are then%
\begin{subequations}
\label{eq:Eckart-frame-1}
\begin{align}
   & T^{\mu\nu} = \epsilon u^\mu u^\nu + \left[ p +f_1 \dot T/T + f_2 \partial{\cdot}u + f_3 u^\lambda \partial_\lambda(\mu/T) \right] \Delta^{\mu\nu} \nonumber\\
  & \ \ \ \ \ + ({\cal Q}^\mu u^\nu + {\cal Q}^\nu u^\mu) - \eta \sigma^{\mu\nu} + O(\partial^2)\,,\\
  & J^\mu  = n u^\mu +  O(\partial^2)\,,
\end{align}
\end{subequations}
with 
\begin{align}
\label{eq:QE}
  {\cal Q}^\mu = -\frac{\epsilon+p}{n} \left(\ell_1  \dot u^\mu + \ell_2/T \,\Delta^{\mu\lambda} \partial_\lambda T + \ell_3 \Delta^{\mu\lambda} \partial_\lambda (\mu/T) \right) + O(\partial^2)\,.
\end{align}
The conditions ${\cal E} = \epsilon$, ${\cal N} = n$, and ${\cal J}^\mu=0$ are usually taken as the defining properties of the Eckart frame.
It is clear that the Eckart frame can only be defined for charged fluids, and then it is only useful when one studies states with non-zero charge density $n$. Again, there are multiple ways to write down the on-shell Eckart-frame constitutive relations, by using the on-shell relations among the one-derivative scalars and vectors. For example, eliminating $\Delta^{\mu\lambda} \partial_\lambda (\mu/T)$ from (\ref{eq:QE}) one finds
\begin{align}
\label{eq:QE-1}
  {\cal Q}^\alpha = -\kappa\left( T\dot u^\alpha + \Delta^{\alpha\lambda} \partial_\lambda T\right) + \chi_{\rm T} \Delta^{\alpha\lambda} \partial_\lambda T + O(\partial^2)\,.
\end{align}
Here $\kappa \equiv (\epsilon+p)^2 \sigma/(n^2 T)$ is the heat conductivity which is proportional to the charge conductivity $\sigma$. Again, the coefficient $\chi_{\rm T}$ must vanish by the thermodynamic consistency conditions. The constitutive relations in the form (\ref{eq:QE-1}) were proposed in the original paper~\cite{PhysRev.58.919} by Eckart%
\footnote{
Note that the discussion of Ohm's law in \cite{PhysRev.58.919} is incomplete. The correct form of the Ohm's law in relativistic fluids is not $J^\alpha = \sigma E^\alpha$, but rather $J^\alpha = \sigma (E^\alpha - T \Delta^{\alpha\nu}\partial_\nu(\mu/T))$, as is required by the thermodynamic consistency conditions.
}.
Alternatively, eliminating $\dot u^\mu$ from (\ref{eq:QE}) one finds
\begin{align}
\label{eq:QE-2}
  {\cal Q}^\alpha =\frac{\epsilon{+}p}{n}\sigma T \Delta^{\alpha\lambda}\partial_\alpha(\mu/T) - \frac{\epsilon{+}p}{n}\chi_{\rm T}\Delta^{\lambda\alpha}\partial_\alpha T + O(\partial^2)\,.
\end{align}
Two different Eckart-frame constitutive relations (\ref{eq:QE-1}) and (\ref{eq:QE-2}) will give rise to inequivalent hydrodynamic equations in one-derivative hydrodynamics. In particular, the modes that are stable with the constitutive relations (\ref{eq:QE-2}) become unstable with the constitutive relations (\ref{eq:QE-1}), see e.g.~\cite{Kovtun:2012rj}. Of course, there is also a similar freedom to redefine the non-equilibrium contributions to pressure, for example by shifting them from $\partial{\cdot}u$ to $\dot T$.

\section{Stability of equilibrium}
\label{sec:stability}
In the rest of this paper we will study the stability of small perturbations of the thermal equilibrium state in flat space with no external electromagnetic fields. The equilibrium state is characterized by the temperature $T$, fluid velocity $u^\alpha = (1,\v)/\sqrt{1-\v^2}$, and (for charged fluids) the chemical potential $\mu$. In order to study the linearized stability, we take $T=T_0 + \delta T$, $\v = \v_0 +\delta\v$, $\mu = \mu_0 +\delta\mu$, and linearize the hydrodynamic equations in $\delta T$, $\delta\v$, $\delta\mu$, with constant $T_0$, $\mu_0$ and $\v_0$, such that $|\v_0|<1$. The solutions to the linearized equations may be taken as combinations of plane waves, and we take $\delta T$, $\delta\v$, and $\delta\mu$ proportional to $e^{i\k{\cdot}\x - i\omega t}$. Solving the hydrodynamic equations $\partial_\mu T^{\mu\nu} = 0$, $\partial_\mu J^\mu = 0$ with the general constitutive relations (\ref{eq:EPQTNJ}), one finds the set of eigenfrequencies $\omega(\k)$. The eigenfrequencies depend on $T_0$, $\v_0$ and $\mu_0$, as well as on all the transport coefficients in~(\ref{eq:EPQTNJ}). By ``stability'' we will mean the following:
\begin{align}
\label{eq:Imwl0}
   {\rm Im}\ \omega(\k) \leqslant 0\,,
\end{align}
for all $\k$. This ensures that small perturbations of the equilibrium state do not grow exponentially with time. By ``causality'' we will mean the following:
\begin{align}
\label{eq:cc}
  \lim_{k\to\infty} \frac{{\rm Re}\;\omega(\k)}{k} < 1\,,
\end{align}
where $k\equiv |\k|$. See e.g.~\cite{Krotscheck1978} for causality criteria, though the general discussion of causality and hyperbolicity is beyond the scope of the present paper.%
\footnote{
  In ${\cal N}=4$ supersymmetric, $SU(N_c)$ Yang-Mills theory at strong coupling, exact dispersion relations at large $N_c$ can be found for all modes by using the holographic duality, without using the hydrodynamic approximation. The dispersion relations are such that, as $k$ increases, ${\rm Re}\,\omega(k)/k$ approaches one from above~\cite{Kovtun:2005ev,Festuccia:2008zx,Fuini:2016qsc}. Of course, in this case the short-distance behaviour is regulated by the actual microscopic degrees of freedom of the quantum field theory, instead of the modes contained in the hydrodynamic equations. 
}
As we will see later, if the causality condition (\ref{eq:cc}) is not satisfied for the fluid at rest, then the uniformly moving fluid will have unstable modes. This follows from Lorentz covariance of the hydrodynamic equations.

The algebraic equation that determines the eigenfrequencies is obtained by setting the determinant of the corresponding linear system to zero. In a relativistic fluid, for an equilibrium state specified by a constant velocity $\bar u^\alpha$, the algebraic equation can only depend on $q{\cdot}\bar u$ and $q^2$, where $q^\alpha=(\omega,\k)$. Alternatively, we can take the equation to depend on $w\equiv -q{\cdot}\bar u$ and $z\equiv q^2 + (q{\cdot}\bar u)^2 = q_\alpha \bar\Delta^{\alpha\beta} q_\beta$, so that the eigenfrequencies are determined by
\begin{align}
\label{eq:H1}
  H(w,z) = 0\,.
\end{align}
In a rotation-invariant fluid, the modes split into the ``shear-channel'' (perturbations of $\delta\v \perp \k$) and ``sound-channel'' (perturbations of $\delta\v \parallel\k$, $\delta T$, $\delta\mu$) modes.%
\footnote{
  For charged fluids at $\k\to0$, the sound channel fluctuations further split into the proper sound modes with $\omega = \pm \vs |\k| + \dots$ and the heat diffusion mode with $\omega = -iD\k^2 + \dots$.
}
This is readily seen for the fluid at rest, with $\bar u^\alpha=(1,{\bf 0})$, so that $H(\omega,\k^2) = F_{\rm shear}^{d-1}(\omega,\k^2) F_{\rm sound}(\omega,\k^2)$ in $d$ spatial dimensions. By Lorentz covariance, we then have
\begin{align}
\label{eq:H2}
  H(w,z) = F_{\rm shear}^{d-1}(w,z) F_{\rm sound}(w,z)\,.
\end{align}
The stability can then be analyzed independently in the shear channel and in the sound channel. To any finite order in the derivative expansion, the functions $F_{\rm shear}(w,z)$, $F_{\rm sound}(w,z)$ are finite-order polynomials in $w$ and $z$.

Consider first-order hydrodynamics of uncharged fluids in the general frame (\ref{eq:EPQTNJ}). Thermodynamic consistency conditions imply $\theta_1 = \theta_2$, thus in a general frame in one-derivative hydrodynamics we have six transport coefficients: $\varepsilon_{1,2}$, $\pi_{1,2}$, $\theta\equiv \theta_1=\theta_2$, and $\eta$. As we will see, in this six-dimensional parameter space of transport coefficients, there is a subspace where Eq.~(\ref{eq:Imwl0}) is true. This subspace defines a class of stable frames in uncharged fluids. 

For charged fluids in the general frame (\ref{eq:EPQTNJ}), we further have $\gamma_1=\gamma_2$ by the thermodynamic consistency conditions, thus there are fourteen transport coefficients in one-derivative hydrodynamics: $\varepsilon_{1,2,3}$, $\pi_{1,2,3}$, $\nu_{1,2,3}$, $\theta\equiv \theta_1=\theta_2$, $\theta_3$, $\gamma\equiv\gamma_1=\gamma_2$, $\gamma_3$, $\eta$. Again, we expect that in this fourteen-dimensional parameter space of transport coefficients, there is a subspace where Eq.~(\ref{eq:Imwl0}) is true. This subspace defines a class of stable frames for charged fluids.

Among all the transport coefficients, we expect that the ones that come with time derivatives will be more important for ensuring stability than the rest. We thus expect that beyond the standard transport coefficients $\eta>0,\zeta>0,\sigma>0$, the coefficients that are most relevant for the stability of one-derivative hydrodynamics are $\varepsilon_1$, $\pi_1$, $\theta_1$, $\nu_1$, and $\gamma_1$. These coefficients can be thought of as defining  relaxation times for the energy density ($\varepsilon_1$), pressure ($\pi_1$), momentum density ($\theta_1$), charge density ($\nu_1$), and the charge current ($\gamma_1$).

\section{Uncharged fluids in the general frame}
\label{sec:uncharged}
The linearized fluctuations of the equilibrium state split into the ``shear channel'' and ``sound channel'' modes. At $\v_0=0$, the shear channel modes are the velocity fluctuations with $\delta\v \perp \k$, while the sound channel modes are the coupled fluctuations of $\delta\v \parallel\k$ and $\delta T$. Upon solving the linearized equations $\partial_\mu T^{\mu\nu}=0$, the eigenfrequencies $\omega(\k)$ are found by setting the determinant of the corresponding linear system to zero. The vanishing of the determinant can be written as $F_{\rm shear}(\omega,\k)^{d-1} F_{\rm sound}(\omega,\k) = 0$, corresponding to the two channels. In first-order hydrodynamics, $F_{\rm shear}(\omega,\k)$ is a second-order polynomial in $\omega$ whose coefficients depend on $\k$, while $F_{\rm sound}(\omega,\k)$ is a fourth-order polynomial in $\omega$ whose coefficients depend on $\k$.

The values of $F(\omega,\k)$ at different $\v_0$ are related by boost covariance, according to Eq.~(\ref{eq:H2}). To be specific, let us choose coordinates such that the plane spanned by $\v_0$ and $\k$ is the $xy$-plane, and $\v_0$ is along $x$. Then the equation $F_a(\v_0,\omega',\k')=0$ is equivalent to
\begin{equation}
\label{eq:FF-boosted}
  F_a\left( \!\v_0 {=} 0,\ \omega = \frac{\omega' - k'_x v_0}{\sqrt{1-v_0^2}},\ k_x = \frac{k'_x - \omega' v_0}{\sqrt{1-v_0^2}}, k_y = k'_y \right) = 0\,,
\end{equation}
where ``$a$'' stands for either shear or sound. Note that, in general, the above boost covariance does not imply any simple algebraic relations between $\omega'(\k')$ and $\omega(\k)$. In particular, if for certain values of the transport coefficients we find ${\rm Im}\ \omega(\k)<0$ for the fluid at rest, this does not guarantee that ${\rm Im}\ \omega'(\k')<0$ in a moving fluid (the modes that are stable at $\v_0{=}0$ can become unstable at $\v_0{\neq}0$). Thus, the analysis of stability has to be performed at non-zero $\v_0$.

\begin{figure}[t]
\centering
\includegraphics[width=0.55\textwidth]{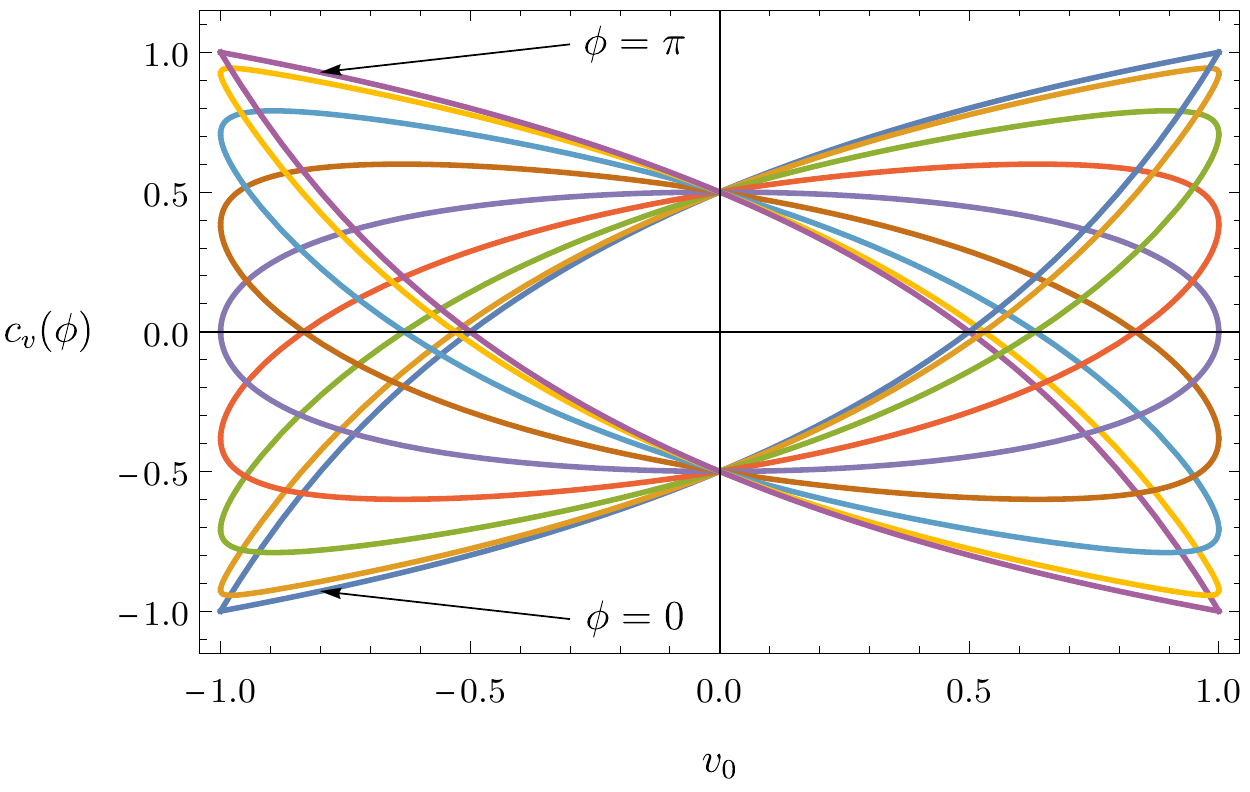}
\caption{
\label{fig:cvphi}
Phase velocity $c_v(\phi)$ for waves with a linear dispersion relation in a moving fluid, shown for different angles $0{\leqslant}\phi{\leqslant}\pi$ of the wave vector with respect to $\v_0$. Each colour corresponds to a given value of $\phi$. In the plot, the wave speed in the fluid at rest is taken $c_0=\frac12$.
}
\end{figure}

A simplification occurs if the dispersion relation at $\v_0=0$ happens to be linear. This is true for sound waves at small $\k$, and happens more generally for all modes at large $\k$. Let $\omega(\k) = \pm c_0 |\k|$, with some constant speed $0<c_0<1$. Performing a Lorentz boost on $(\omega,\k)$ as in Eq.~(\ref{eq:FF-boosted}) gives again a linear dispersion relation at $v_0\neq0$, namely $\omega' = c_v(\phi)|\k'|$, where $\phi$ is the angle between $\k'$ and $\v_0$. Explicitly, we find
\begin{align}
\label{eq:cvphi}
   c_v(\phi) = \frac{v_0(1{-}c_0^2)}{1{-}c_0^2 v_0^2}\cos\phi \pm \frac{c_0}{(1{-}c_0^2 v_0^2)} \sqrt{(1{-}v_0^2) \left[1 - v_0^2 c_0^2 - v_0^2 (1{-}c_0^2)\cos^2\phi \right]}\,.
\end{align}
At $\phi=0$ the above reduce to $c_v = (v_0 \pm c_0)/(1\pm v_0 c_0)$, the standard relativistic addition of phase velocities. Figure~\ref{fig:cvphi} illustrates how the wave velocity $c_v(\phi)$ in a moving fluid depends on $v_0$ and $\phi$. As expected, $c_v(\phi)^2<1$ for $v_0^2<1$. 

For sound waves, Eq.~(\ref{eq:cvphi}) gives the sound velocity in a moving relativistic fluid, with $c_0=(\partial p/\partial\epsilon)^{1/2}$. For large-momentum modes, Eq.~(\ref{eq:cvphi}) implies that if the modes at $v_0=0$ are stable ($c_0$ is real) and causal ($0<c_0<1$), they remain so at $v_0\neq0$. Alternatively, $c_0>1$ for the fluid at rest implies that the equilibrium state of a uniformly moving fluid is unstable.

More generally, being consequences of Lorentz covariance, Eqs.~(\ref{eq:FF-boosted}) and (\ref{eq:cvphi}) are of course not restricted to first-order hydrodynamics. In the context of hydrodynamics, Eq.~(\ref{eq:FF-boosted}) is true to all orders in the derivative expansion.

\subsection{Shear channel}
In the shear channel, the eigenfrequencies are determined by the zeroes of
\begin{align}
\label{eq:shear-quadratic}
  F_{\rm shear} = (\theta - \v_0^2 \eta)\omega^2 + \left( \frac{i w_0}{\gamma_0} - 2(\theta{-}\eta)\k{\cdot}\v_0\right)\omega - \frac{i w_0}{\gamma_0} (\k{\cdot}\v_0) - \frac{\k^2\eta}{\gamma_0^2} + (\k{\cdot}\v_0)^2 (\theta{-}\eta)\,,
\end{align}
where $\gamma_0\equiv 1/\sqrt{1-\v_0^2}$. If we measure $\omega$ and $\k$ in units of $(\epsilon_0{+}p_0)/\eta$, the stability of the shear-channel eigenmodes is determined by only one dimensionless parameters $\theta/\eta$.
There are two gapless ($\omega(\k{\to}0)=0$) modes with different polarizations, corresponding to the familiar shear waves, and two gapped ($\omega(\k{\to}0)\neq0$) modes with different polarizations. For the eigenfrequencies, we find at long wavelength
\begin{align}
\label{eq:wshear-gapless}
  & \omega(\k) = \k{\cdot}\v_0 - \frac{i\eta}{w_0}\sqrt{1-\v_0^2} \left( \k^2 - (\k{\cdot}\v_0)^2 \right) + O(\k^3)\,,\\[5pt]
\label{eq:wshear-gapped}
  & \omega(\k) = \frac{iw_0\sqrt{1-\v_0^2}}{\eta \v_0^2 - \theta} + O(\k{\cdot}\v_0)\,,
\end{align}
where $w_0\equiv\epsilon_0 + p_0$ is the equilibrium density of enthalpy. It is clear that stability (${\rm Im}\,\omega <0$) of the shear channel fluctuations requires
\begin{align}
\label{eq:shear-stability}
  \theta > \eta > 0\,.
\end{align}
The Landau-Lifshitz convention sets $\theta=0$ at non-zero $\eta$, hence the Landau-Lifshitz hydrodynamics predicts that the thermal equilibrium state is unstable~\cite{Hiscock:1985zz}. The gapped mode which is responsible for the instability is outside of the validity regime of the hydrodynamic approximation, and therefore its physical predictions should not be trusted. Nevertheless, it is important that ${\rm Im}\,\omega(\k) <0$ is true for all modes, both gapless and gapped, in order to ensure that the hydrodynamic equations have predictive power at macroscopic times.

\begin{figure}
\includegraphics[width=0.48\textwidth]{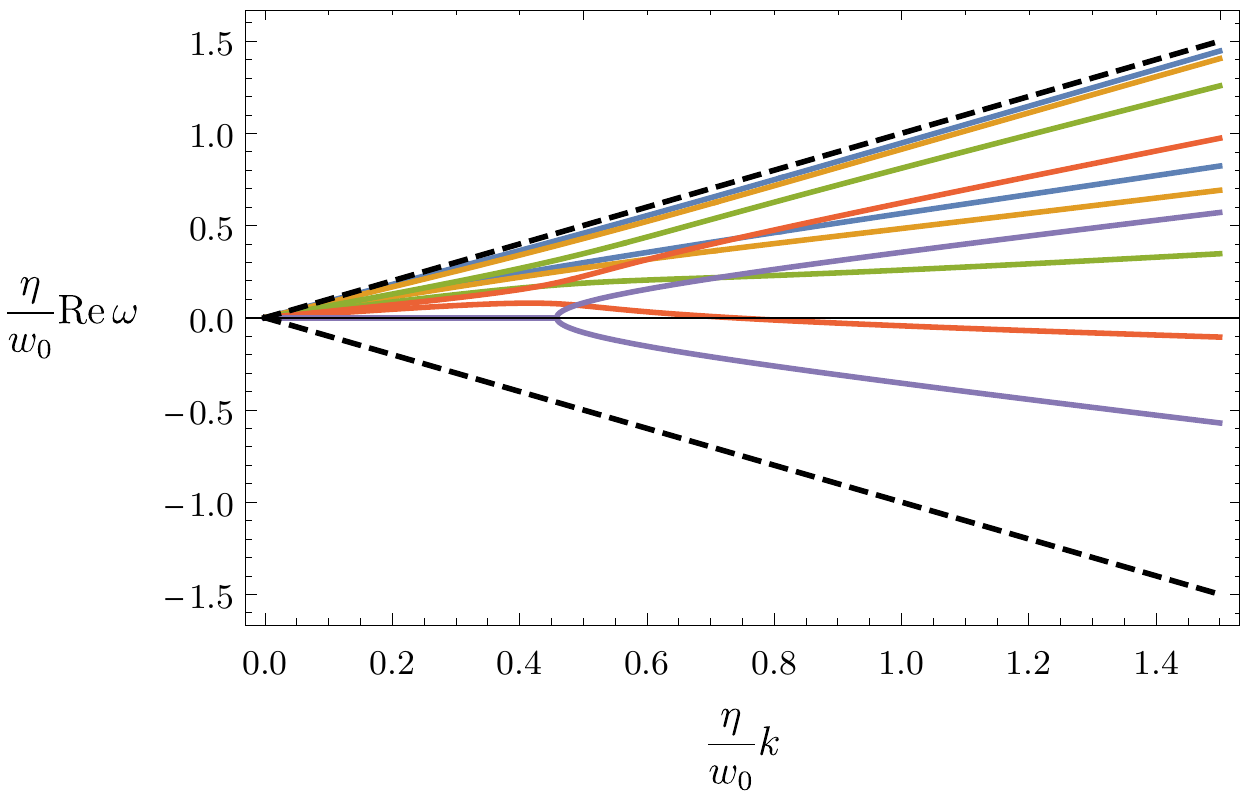}
\hspace{0.02\textwidth}
\includegraphics[width=0.48\textwidth]{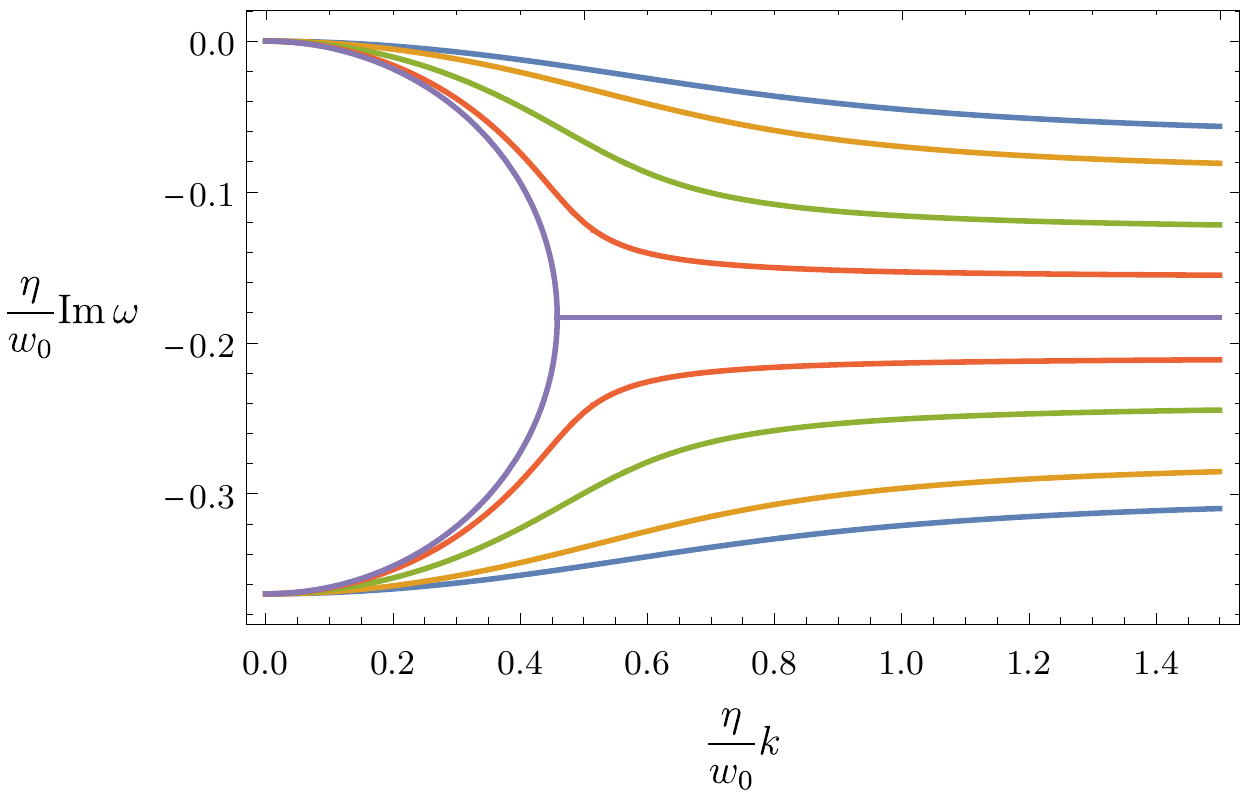}
\caption{
\label{fig:shear-disp-plots}
Real and imaginary parts of the shear channel eigenfrequencies, shown for $v_0=0.9$ and $\theta/\eta=2$, for different angles $0\leqslant\phi\leqslant\pi/2$ of the wave vector $\k$ with respect to $\v_0$. Each colour corresponds to a given value of $\phi$ (blue corresponds to $\phi=0$, purple to $\phi=\pi/2$), $k\equiv |\k|$, and $w_0\equiv(\epsilon_0{+}p_0)$. At small $k$, the gapless and the gapped modes (\ref{eq:wshear-gapless}), (\ref{eq:wshear-gapped}) are clearly visible. At large $k$, the modes follow a linear dispersion relation $\omega = c_{\rm shear}(\phi)k$, with the velocity $c_{\rm shear}(\phi)$ determined by Eq.~(\ref{eq:cs-shear}). The dashed lines denote the light cone $\omega=\pm k$.
}
\end{figure}
At large $\k$, the modes follow a linear dispersion relation, $\omega = c_{\rm shear}(\phi)k$, where $\phi$ is the angle between $\v_0$ and $\k$. The velocity $c_{\rm shear}$ determined by
\begin{equation}
\label{eq:cs-shear}
 \left(\theta {-} v_0^2 \eta \right) c_{\rm shear}^2 - 2v_0\cos\phi \left(\theta {-} \eta \right) c_{\rm shear} + v_0^2 \left(\theta \cos^2\phi + \eta\sin^2\phi \right) - \eta = 0\,,
\end{equation}
with $|v_0|<1$. The solutions are real and bounded by 
$$
  |c_{\rm shear}| < \frac{1+|v_0|(\theta/\eta)^{1/2}}{|v_0|+ (\theta/\eta)^{1/2}}<1\,,
$$
for $\theta>\eta$. As an illustration, the dispersion relations in the shear channel are plotted in Fig.~\ref{fig:shear-disp-plots}, with $\omega$ and $|\k|$ in units of $(\epsilon_0{+}p_0)/\eta$.

A short exercise shows that the condition (\ref{eq:shear-stability}) guarantees that the solutions of (\ref{eq:shear-quadratic}) have ${\rm Im}\, \omega(\k)\leqslant 0$ for all $\k$. Thus Eq.~(\ref{eq:shear-stability}) is the stability criterion for shear-channel fluctuations in first-order relativistic hydrodynamics.

\subsection{Sound channel}
We now come to sound-channel oscillations. In the sound channel, the eigenfrequencies $\omega(\k)$ are determined by the zeroes of
\begin{align}
\label{eq:Fsound0}
  F_{\rm sound}(\v_0{=}0,\omega,\k) 
  & = \vs^2 \varepsilon_1 \theta\, \omega^4
      + i w_0 (\vs^2 \varepsilon_1 {+} \theta)\omega^3 
      -i \k^2 w_0 \left(\gamma_s + \vs^4\varepsilon_1 + \vs^2\theta \right)\omega \nonumber\\
  & - \left(w_0^2 + \k^2 \vs^2\left(\vs^4\varepsilon_1^2 + \gamma_s \varepsilon_1 +(\varepsilon_2{+}\pi_1)(\theta{-}\vs^2\varepsilon_1) + \varepsilon_2\pi_1 \right) \right)\omega^2 \nonumber\\
  & +\k^2\vs^2 \left(w_0^2 + \k^2\theta (\vs^2(\varepsilon_2 {+} \pi_1 {-} \vs^2\varepsilon_1) - \gamma_s) \right)\,, 
\end{align}
where $\vs^2\equiv\partial p_0/\partial\epsilon_0$ is the speed of sound, $\gamma_s\equiv \frac43\eta + \zeta$ sets the damping of sound waves, and $w_0\equiv \epsilon_0+p_0$ as before.%
\footnote{
The coefficient $\pi_2$ has been traded for the bulk viscosity $\zeta$, according to the formula in footnote~\ref{fn:zeta}. 
} 
More generally, $F_{\rm sound}(\v_0{\neq}0)$ can be obtained by Lorentz boosting the four-vector $(\omega,\k)$ in the above $F_{\rm sound}(\v_0{=}0)$, as in Eq.~(\ref{eq:FF-boosted}). If we measure $\omega$ and $\k$ in units of $w_0/\gamma_s$, the stability of the sound-channel eigenmodes is determined by only four dimensionless parameters: $\varepsilon_{1,2}/\gamma_s$, $\pi_1/\gamma_s$, and $\theta/\gamma_s$. We will assume that the equation of state is such that the speed of sound is less than the speed of light, $0<\vs<1$.

\begin{figure}
\includegraphics[width=0.48\textwidth]{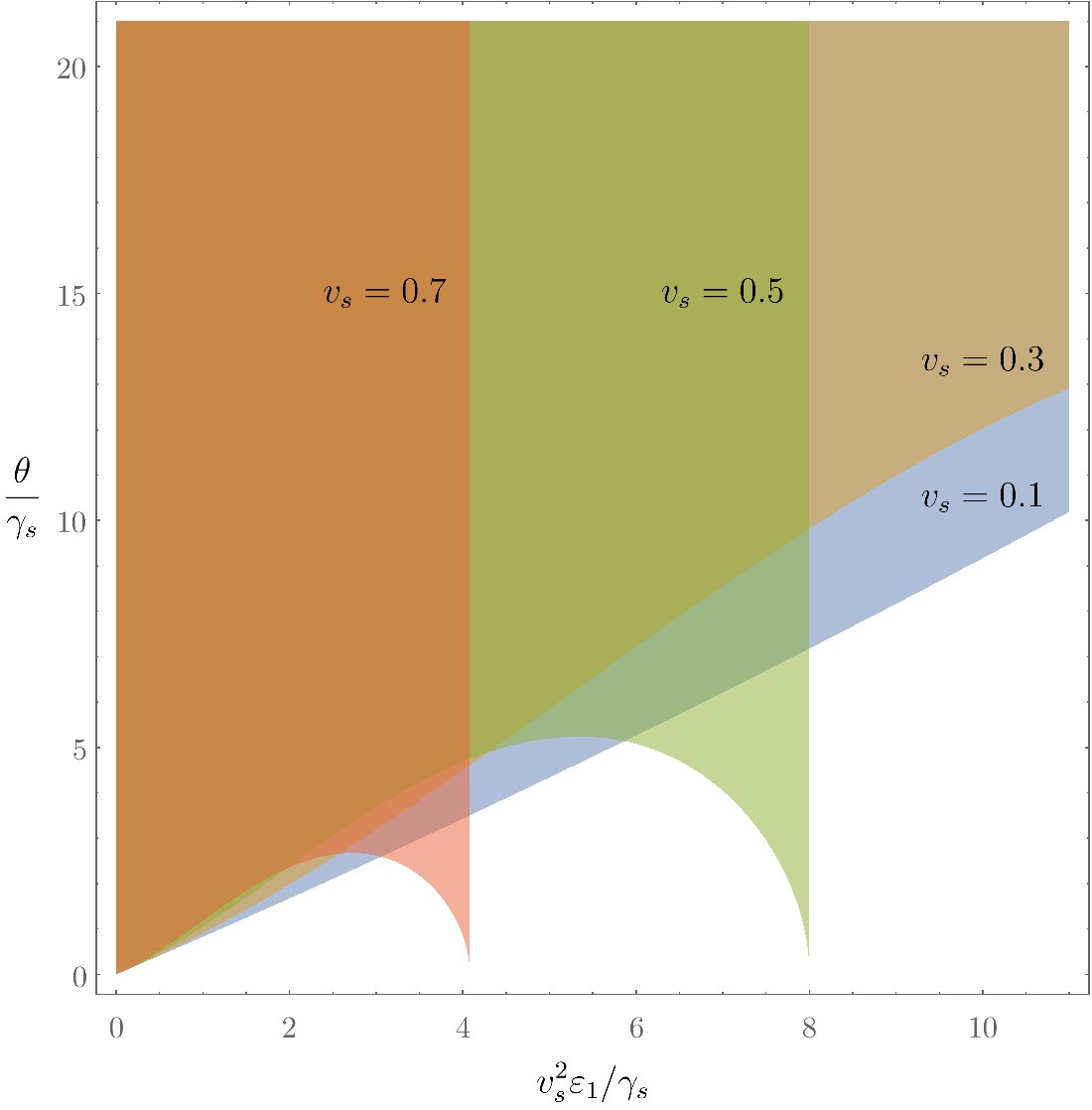}
\hspace{0.02\textwidth}
\includegraphics[width=0.48\textwidth]{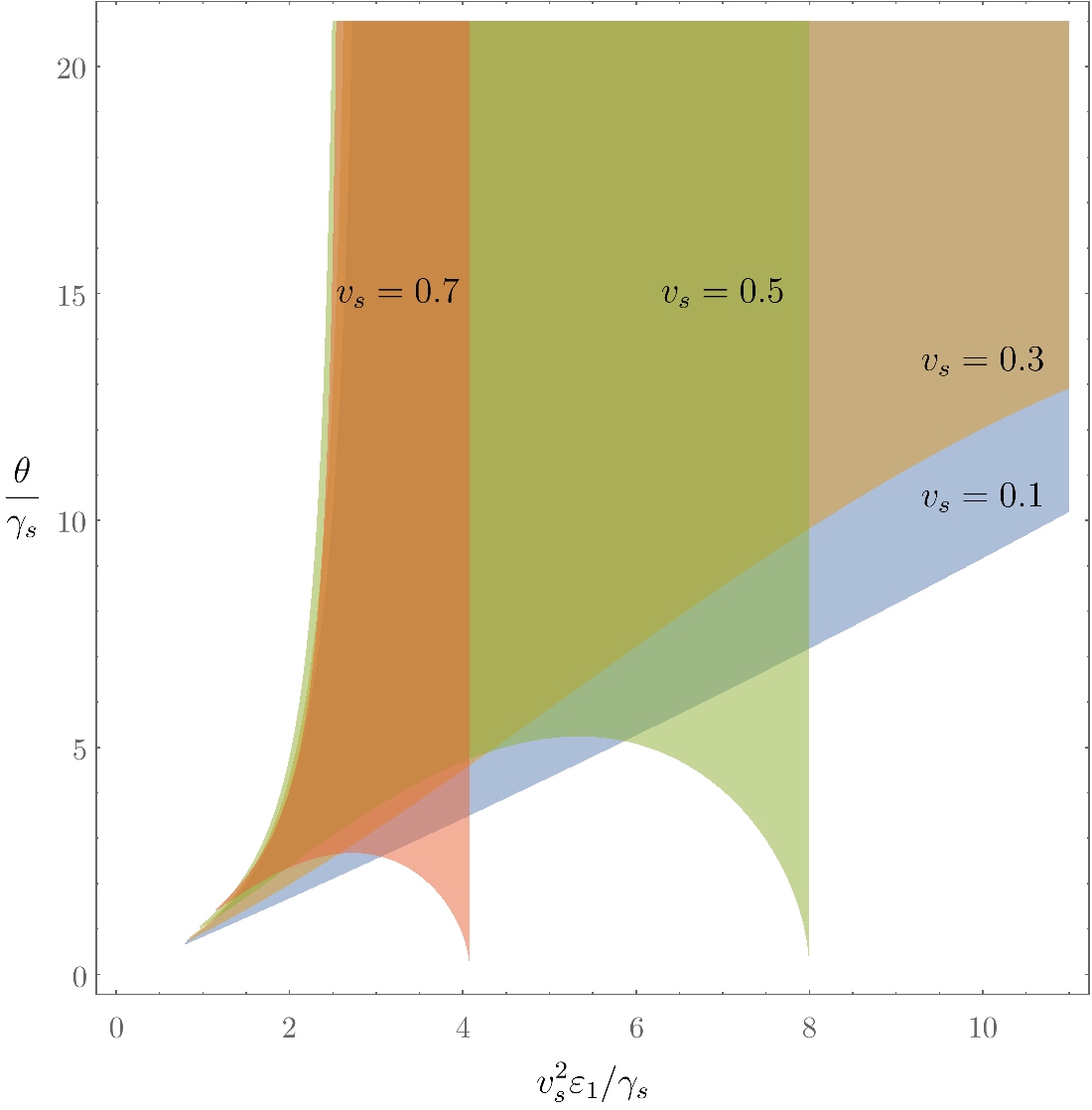}
\caption{
\label{fig:soundstability-1}
Constraints on the transport coefficients $\theta$ and $\varepsilon_1$ obtained by demanding that the sound-channel modes for the fluid at rest are stable and causal. For illustrative purposes, we have taken $\varepsilon_2=0$, $\pi_1/\gamma_s = 3/\vs^2$. The stability region is shaded with a colour corresponding to a given value of $\vs$. The stability region is larger for smaller $\vs$. Left: the region where all modes are stable.  Right: the region where all modes are stable and the short-wavelength modes are causal, $\lim_{k\to\infty}|\omega(k)/k|<1$. In the right plot, the origin $\varepsilon_1=\theta=0$ is always outside the stability region.
}
\end{figure}

To get a sense of the sound-channel stability constraints, let us look at the fluid at rest first, i.e.\ $\v_0=0$. At small $\k$, there are two sound waves, and two gapped modes:
\begin{align}
  & \omega(\k) = \pm \vs |\k| - \frac{i}{2}\frac{\gamma_s}{\epsilon_0 {+} p_0} \k^2 + O(k^3)\,,\\
  & \omega(\k) = -i\frac{\epsilon_0{+}p_0}{\vs^2 \varepsilon_1} + O(k^2)\,,\ \ \ \ 
    \omega(\k) = -i\frac{\epsilon_0{+}p_0}{\theta} + O(k^2)\,.
\label{eq:wgaps-2}
\end{align}
Clearly, stability in the sound channel requires 
\begin{align}
  \gamma_s>0\,,\ \ \varepsilon_1>0\,,\ \ \theta>0\,.
\end{align}

At arbitrary $\k$, we have to study the zeroes of (\ref{eq:Fsound0}). Setting $\omega=i\Delta$ in Eq.~(\ref{eq:Fsound0}) gives rise to a quartic equation for $\Delta$ with real coefficients, which can be written as 
\begin{equation}
\label{eq:Delta-eqn-1}
  a\Delta^4 + b \Delta^3 + c\Delta^2 + d\Delta + e = 0\,,
\end{equation}
such that $a>0$, $b>0$, $d>0$ and the coefficients $a,b,c,d,e$ can be read off by comparing (\ref{eq:Delta-eqn-1}) and (\ref{eq:Fsound0}). Stability requires ${\rm Re}(\Delta) <0$, which by the Routh-Hurwitz criterion amounts to $e>0$ and $\frac{e}{a} < \frac{d}{b}(\frac{c}{a} - \frac{d}{b})$. The first of these conditions gives
\begin{equation}
\label{eq:sound-c1}
  \varepsilon_2 + \pi_1 > \frac{\gamma_s}{\vs^2} + \vs^2 \varepsilon_1\,,
\end{equation}
while the second gives an inequality which involves $\varepsilon_{1,2}$, $\pi_1$, and $\theta$ in a non-linear way. The latter can be written as
\begin{equation}
\label{eq:sound-c2}
  \frac{\bar\varepsilon_1^2}{\vs^2} + \vs^2 (\bar\varepsilon_1 {-} \bar\varepsilon_2) (\bar\varepsilon_1 {+} \bar\theta)^2 (\bar\varepsilon_1 {-} \bar\pi_1) + (\bar\varepsilon_1 {+} \bar\theta) \left(2\bar\varepsilon_1^2 -\bar\varepsilon_1 (\bar\varepsilon_2 {+} \bar\pi_1) + (\bar\theta {+} \bar\varepsilon_2)(\bar\theta {+} \bar\pi_1) \right) >0,
\end{equation}
where $\bar\varepsilon_1 \equiv \vs^2 \varepsilon_1/\gamma_s$, $\bar\varepsilon_2 \equiv \varepsilon_2/\gamma_s$, $\bar\theta \equiv \theta/\gamma_s$, $\bar\pi_1 \equiv \pi_1/\gamma_s$ are dimensionless.
The constraints on the transport coefficients obtained by demanding that the sound-channel modes at $\v_0=0$ are stable are illustrated in Fig.~\ref{fig:soundstability-1}, left.

Finally, we look at large-$k$ modes for the fluid at rest. The modes have a linear dispersion relation, $\omega=c_{\rm sound} k$, with $c_{\rm sound}$ determined by
\begin{align}
\label{eq:cs0largek}
  \frac{\bar\varepsilon_1 \bar\theta}{\vs^2} c_{\rm sound}^4 + 
  \left(\bar\varepsilon_1  (\bar\varepsilon_2 {+} \bar\pi_1 {-} \bar\varepsilon_1 {-}\coeff{1}{\vs^2} ) {-} \bar\theta(\bar\varepsilon_2 {+} \bar\pi_1) {-} \bar\varepsilon_2 \bar\pi_1 \right) c_{\rm sound}^2
  + \bar\theta \left( \vs^2 (\bar\varepsilon_2 {+} \bar\pi_1 {-} \bar\varepsilon_1) {-} 1 \right) = 0.
\end{align}
We want the solutions of this quadratic equation for $c_{\rm sound}^2$ to be real and positive (for stability) and less than one (for causality). The stability constraints from Eq.~(\ref{eq:cs0largek}) are weaker than the general stability constraint of Eqs.~(\ref{eq:sound-c1}), (\ref{eq:sound-c2}) (as expected), but the causality constraint from Eq.~(\ref{eq:cs0largek}) gives something extra. The stability constraints combined with the causality constraint $c_{\rm sound}^2<1$ for the fluid at rest are illustrated in Fig.~\ref{fig:soundstability-1}, right.%
\footnote{
For a quadratic equation of the form $ax^2 + bx+c=0$ with $a>0$ and real $b,c$, demanding that $x$ is real and $0<x<1$ amounts to $b^2>4ac$, $0<c<a$, $-c-a<b<0$. Applying this to Eq.~(\ref{eq:cs0largek}) with $x=c_{\rm sound}^2$ gives explicit causality constraints on the one-derivative transport coefficients.
}

\begin{figure}
\includegraphics[width=0.48\textwidth]{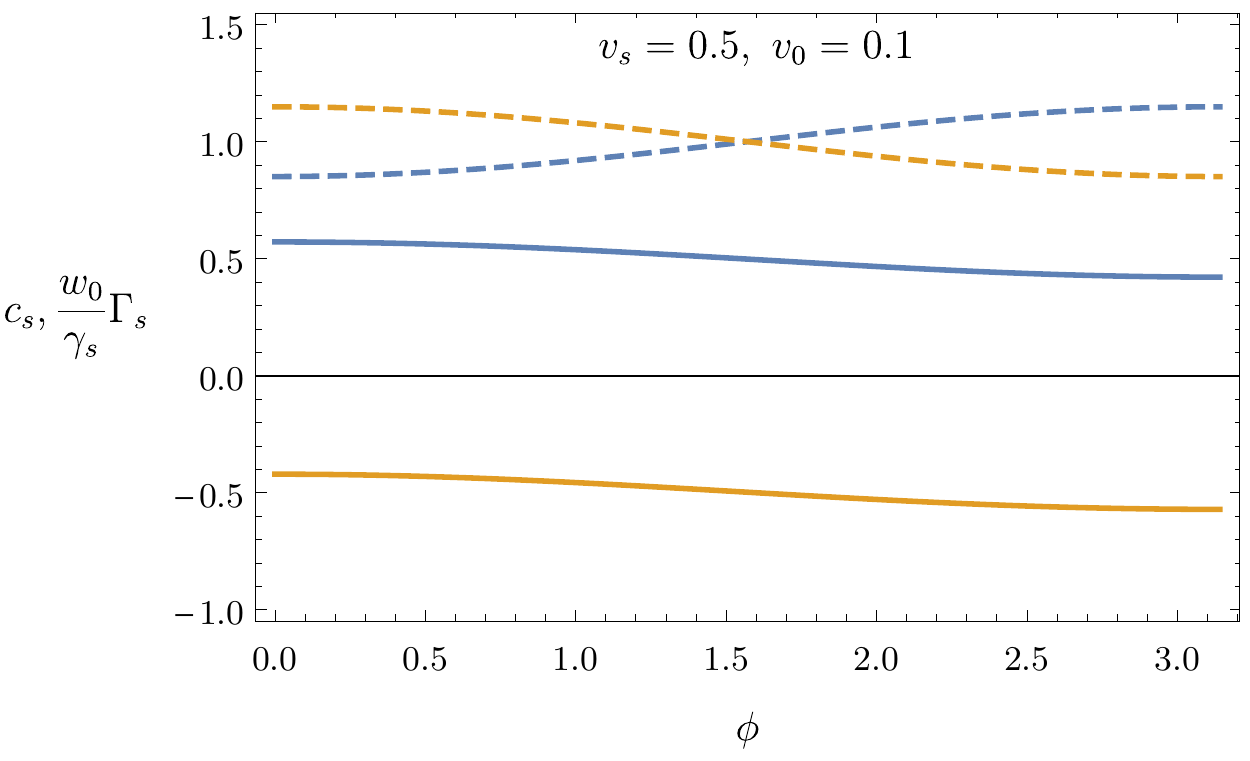}
\hspace{0.02\textwidth}
\includegraphics[width=0.48\textwidth]{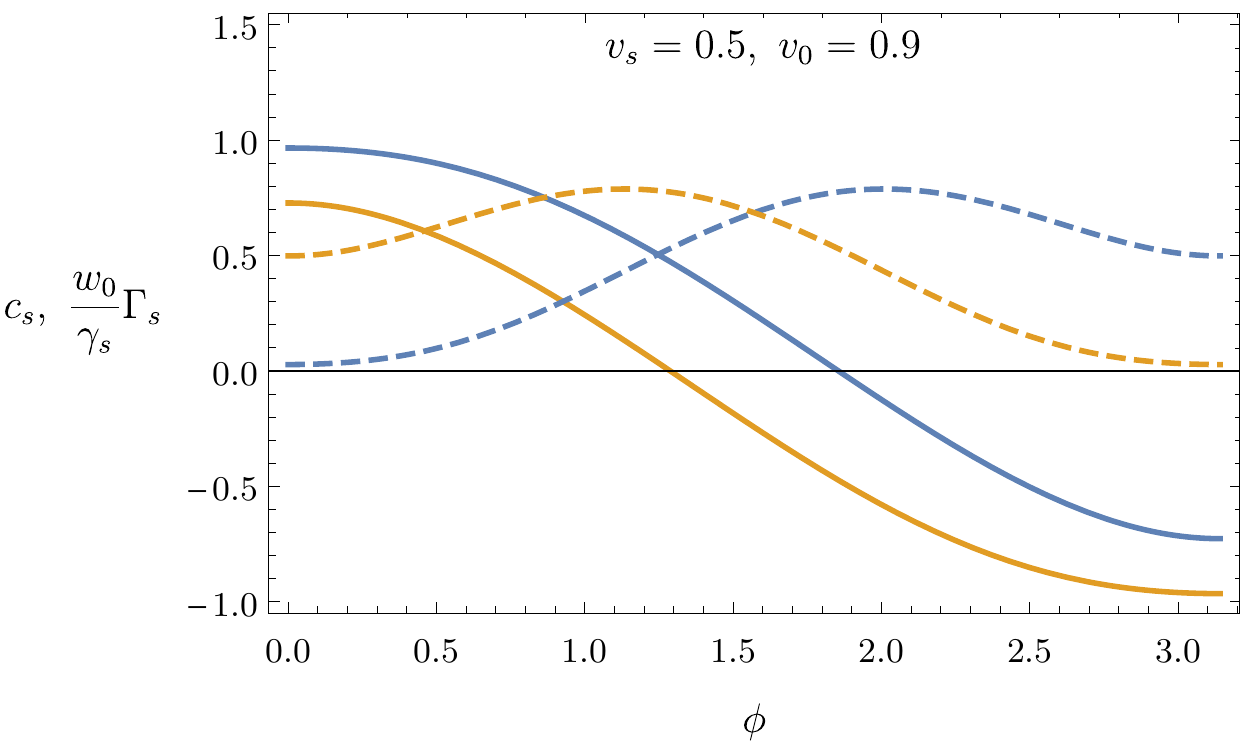}
\caption{
\label{fig:soundspeeds}
Sound velocities $c_s$ and sound damping coefficients $\Gamma_s$ for the two branches of sound in a moving fluid, shown as functions of the angle $\phi$ between $\k$ and $\v_0$. The damping coefficient $\Gamma_s$ is plotted in units of $\gamma_s/(\epsilon_0 {+}p_0)$. Each colour corresponds to a given branch, with $c_s$ shown by solid curves, and $\Gamma_s$ by dashed curves. For illustrative purposes, we have taken the speed of sound for the fluid at rest to be $\vs=1/2$.
}
\end{figure}

Let us now consider sound-channel oscillations in a moving fluid. We start with sound waves (small $k$).
For sound waves that propagate parallel to the fluid flow ($\k\parallel\v_0$), we have
\begin{align}
  \omega(k) = \frac{v_0\pm\vs}{1\pm v_0 v_s} k - \frac{i}{2} \frac{\gamma_s}{\epsilon_0{+}p_0} \frac{(1-v_0^2)^{3/2}}{(1\pm v_0 \vs)^3} k^2 + O(k^3)\,,
\end{align}
where $k\equiv|\k|$, and $v_0$ is between $-1$ and $1$. The sound velocity obeys the standard relativistic velocity addition formula, as follows from the boost covariance condition~(\ref{eq:FF-boosted}). For sound waves that propagate perpendicular to the fluid flow ($\k\perp\v_0$), we have
\begin{align}
  \omega(k) = \pm \frac{(1-v_0^2)^{1/2}}{(1-v_0^2 \vs^2)^{1/2}} \vs k - \frac{i}{2} \frac{\gamma_s}{\epsilon_0{+}p_0} \frac{(1-v_0^2)^{1/2}}{(1 - v_0^2 \vs^2)^2} k^2 + O(k^3)\,.
\end{align}
More generally, for sound waves that propagate at an angle $\phi$ with respect to $\v_0$, we have
\begin{align}
  \omega (\k) = c_s(\phi) k - \frac{i}{2}\Gamma_s(\phi)k^2 + O(k^3)\,.
\end{align}
The sound velocity $c_s(\phi)$ obeys a quadratic equation
\begin{align}
\label{eq:csv-eqn}
  (1{-}v_0^2 \vs^2) c_s^2 - 2v_0\cos\phi (1{-}\vs^2) c_s  +  v_0^2(1{-}\vs^2)\cos^2\phi - (1{-}v_0^2)\vs^2 = 0\,,
\end{align}
whose solutions are real, and $|c_s|<1$. The solutions of Eq.~(\ref{eq:csv-eqn}) are given by (\ref{eq:cvphi}), with $c_0=\vs$, as expected from Lorentz covariance.
Once the sound velocity $c_s(\phi)$ is determined, the damping coefficient is given by
\begin{align}
  \Gamma_s = \frac{\gamma_s}{\epsilon_0 {+} p_0} \frac{c_s {-} v_0\cos \phi}{(1{-}v_0^2)^{1/2}} \frac{1+c_s^2 v_0^2 -2c_s v_0 \cos\phi - v_0^2\sin^2\phi}{ c_s(1{-}v_0^2 \vs^2) - v_0(1{-}\vs^2)\cos\phi }\,.
\end{align}
The angular dependence of $c_s(\phi)$ and $\Gamma_s(\phi)$ is illustrated in Fig.~\ref{fig:soundspeeds} for different values of $v_0$. The damping coefficient $\Gamma_s$ is always positive, as expected.

\begin{figure}
\includegraphics[width=0.48\textwidth]{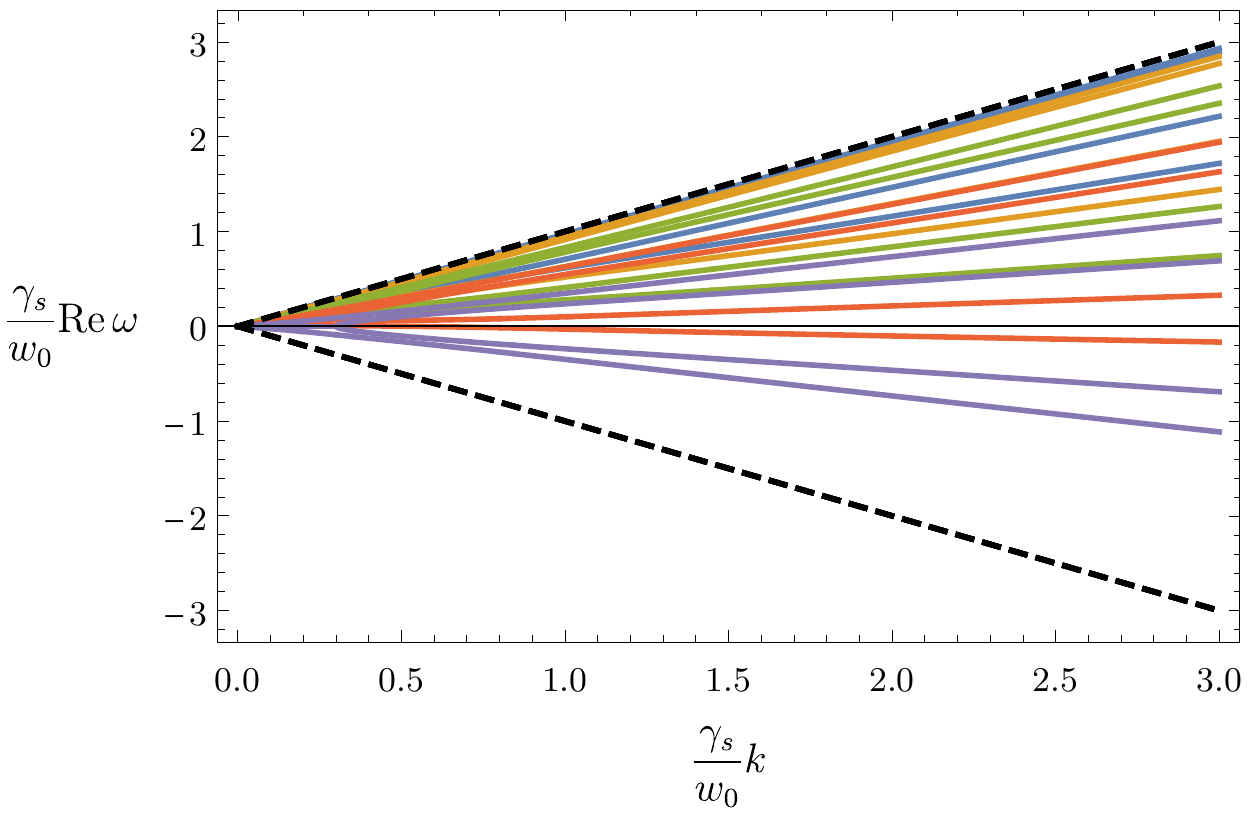}
\hspace{0.02\textwidth}
\includegraphics[width=0.48\textwidth]{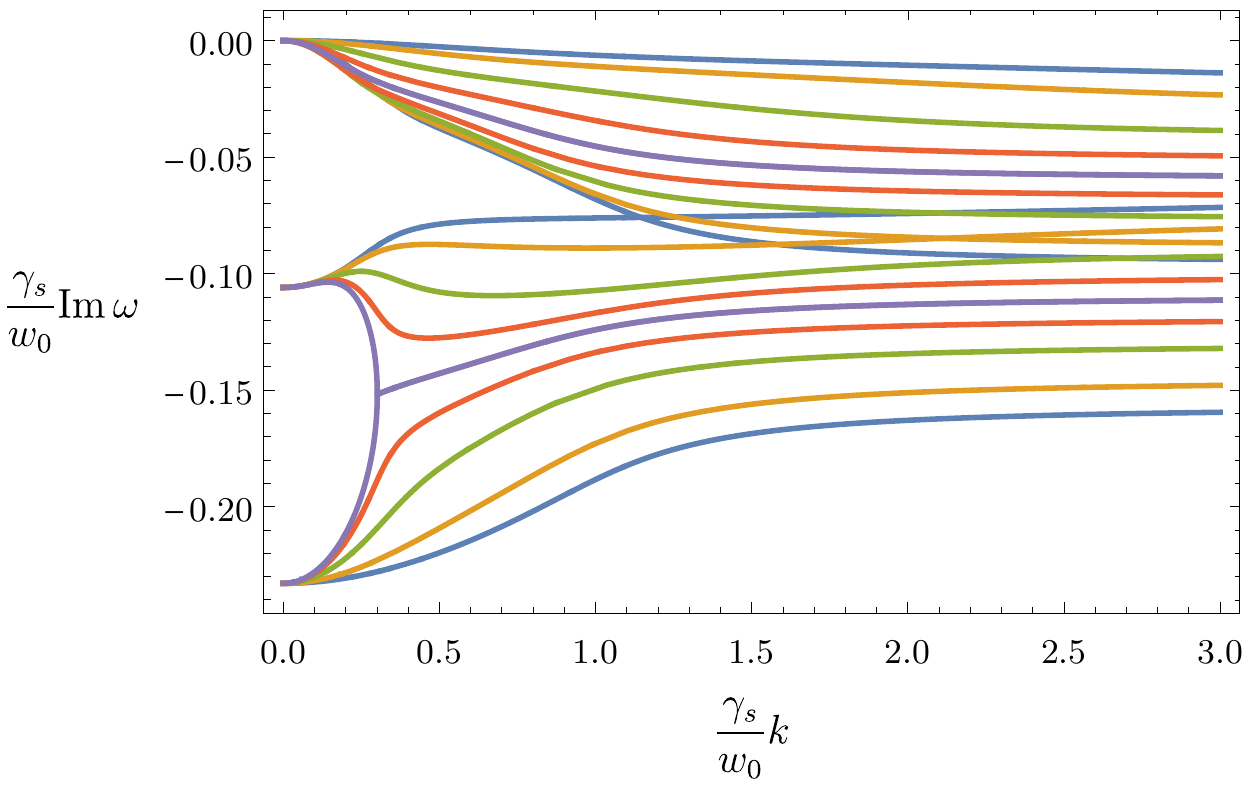}
\caption{
\label{fig:sound-disp-plots}
Real and imaginary parts of the sound channel eigenfrequencies, shown for $v_0=0.9$, and different angles $0\leqslant\phi\leqslant\pi/2$ of the wave vector $\k$ with respect to $\v_0$. Each colour corresponds to a given value of $\phi$ (blue corresponds to $\phi=0$, purple to $\phi=\pi/2$), $k\equiv |\k|$, and $w_0\equiv(\epsilon_0{+}p_0)$. For illustrative purposes, the speed of sound is taken as $\vs=0.5$, and the values of the transport coefficients were taken as follows: $\vs^2\varepsilon_1/\gamma_s = 3$, $\theta/\gamma_s=4$, $\varepsilon_2=0$, $\pi_1/\gamma_s = 3/\vs^2$. The dashed lines denote the light cone $\omega=\pm k$. The values of $\varepsilon_1$ and $\theta$ are inside the stability region of Fig.~\ref{fig:soundstability-1} (right), near the boundary of the stability region.
}
\end{figure}

At non-zero $v_0$, the gapped frequencies (\ref{eq:wgaps-2}) are modified. However, the stability and causality constraints at $v_0=0$ imply that the gaps remain stable at $v_0\neq0$. Further, as mentioned earlier, the stability and causality constraints at $v_0=0$ imply that the stability and causality holds at large $k$ at $v_0\neq0$. Thus the constraints on transport coefficients from the modes at $v_0\neq0$ remain the same as in Fig.~\ref{fig:soundstability-1}, right. The sound-channel dispersion relations are illustrated in Fig.~\ref{fig:sound-disp-plots}. As expected, the dispersion relations are stable and causal when the transport coefficients are inside the stability region.

One of the necessary conditions for the stability of the gaps at $v_0 \neq 0$ is 
\begin{align}
  \vs^2 \varepsilon_1 + \theta > \frac{\gamma_s}{1-\vs^2}\,.
\end{align}
For any finite value of the shear viscosity $\eta{>}0$ this excludes a finite neighbourhood of $\varepsilon_1{=}\theta{=}0$, as indicated in Fig.~\ref{fig:soundstability-1}, right.  The Landau-Lifshitz hydrodynamics of uncharged fluids sets $\varepsilon_1=\theta=0$, and therefore predicts that the thermal equilibrium state at $v_0\neq0$ is unstable.

The stability criteria become particularly simple for hydrodynamics of conformal theories. For uncharged fluids, conformal symmetry in 3+1 dimensions implies that 
\begin{align}
\label{eq:cft-relations}
 \varepsilon_1 = 3\pi_1\,,\ \ \ \ 
 \varepsilon_2 = 3\pi_2\,,\ \ \ \ 
 \pi_1 = 3\pi_2\,,
\end{align}
see Appendix~\ref{sec:conformal}.
The speed of sound is $\vs=1/\sqrt{3}$, and the bulk viscosity vanishes. Thus there are only three independent one-derivative transport coefficients, which can be taken as $\pi_1$, $\theta$, and~$\eta$. The equilibrium state is stable if 
\begin{align}
\label{eq:cft-stable}
  1-\frac{3\eta}{\theta} -\frac{\eta}{\pi_1}>0\,,\ \ \ \ 
  \pi_1 > 4\eta\,.
\end{align}
This agrees with the conditions found in Ref.~\cite{Bemfica:2017wps}. It order to satisfy (\ref{eq:cft-stable}), it is sufficient to take $\theta>4\eta$, $\pi_1>4\eta$ in order to ensure the stability and causality of first-order conformal hydrodynamics.

\section{Discussion}
In this paper we have studied linear perturbations of the thermal equilibrium state in relativistic hydrodynamics. In a Lorentz-covariant theory, if the fluctuations are not causal [in the sense of violating Eq.~(\ref{eq:cc})], they are also unstable. Our focus was on first-order hydrodynamics. The theory is ``first-order'' in the sense that our constitutive relations (\ref{eq:T1}) and~(\ref{eq:EPQTNJ}) contain only terms with up to one derivative of the hydrodynamic variables. The only hydrodynamic variables are the temperature, fluid velocity, and the chemical potential. The first-order constitutive relations in a general ``frame'' contain more transport coefficients than the two viscosities and the heat conductivity. In particular, the first-order constitutive relations contain the relaxation times for the energy density ($\varepsilon_1$), pressure ($\pi_1$), momentum density ($\theta$), charge density ($\nu_1$), and the charge current ($\gamma_1$). 

We have shown that first-order relativistic viscous hydrodynamics can be made linearly stable by a suitable choice of the relaxation times. The minimal first-order regulator for an uncharged fluid is given by only three dimensionless parameters $\theta/\gamma_s$, $\varepsilon_1/\gamma_s$, and $\pi_1/\gamma_s$. The parameters must be bounded from below for stability, in a way illustrated in Fig.~\ref{fig:soundstability-1}, right. Necessary stability conditions include $\theta>\eta>0$, $ \vs^2 \varepsilon_1 + \theta > \gamma_s/(1-\vs^2)$, as well as $ \varepsilon_2 + \pi_1 > (\gamma_s/\vs^2) + \vs^2 \varepsilon_1$. While we haven't undertaken an exhaustive investigation of the multi-dimensional parameter space of transport coefficients, it is clear that there is a finite region in the parameter space where the linear perturbations of the equilibrium state are stable. A more detailed investigation would be straightforward to perform -- the equations that determine the eigenfrequencies $\omega(\k)$ are polynomials in $\omega$, at most fourth order. The stable parameter region in the space of transport coefficients defines a class of frames that are linearly stable. 

We have not performed the stability analysis for charged fluids. Given the constitutive relations (\ref{eq:T1}) and (\ref{eq:EPQTNJ}), doing so would be a straightforward exercise. We expect that the main conclusion about the linear stability of the equilibrium state will still hold, i.e.\ we expect that the parameters $\nu_1$ and $\gamma_1$ will regulate the instabilities of the naive first-order hydrodynamics of charged fluids. 

This paper has an exploratory nature. We view stable first-order frames as ultraviolet regulators of the naive relativistic hydrodynamics, aiming to achieve the same goal as the Israel-Stewart-like approaches. The advantage of stable frames is that no new degrees of freedom are introduced into hydrodynamics. 
The stable one-derivative hydrodynamics is also consistent with the local form of the second law at first order in derivatives, as it should be.
In order to explore the viability of the general-frame first-order hydrodynamics further, it would be interesting to investigate the stability of non-trivial flows, and to see how the first-order hydrodynamics holds up in non-linear evolution.  We hope that the observations of the present paper will lead to further explorations of the general-frame first-order hydrodynamics.

\subsection*{Acknowledgments}
I would like to thank Jorge Noronha for helpful correspondence and Alex Buchel for helpful conversations. This work was supported in part by the NSERC of Canada.

\appendix
\section{Entropy current}
\label{sec:entropy-current}
In this appendix we review the entropy current in first-order hydrodynamics of uncharged fluids (see e.g.\ Ref.~\cite{Bhattacharya:2011tra} for a review, though the discussion below is from a different angle). The canonical entropy current $S^\mu_{\rm canon}$ may be written down using a covariant version of the equilibrium relation $Ts=p+\epsilon$,
\begin{equation}
\label{eq:Smu-def}
  TS^\mu_{\rm canon} = p u^\mu - T^{\mu\nu} u_\nu\,.
\end{equation}
We isolate the ideal part, $T^{\mu\nu} = \epsilon u^\mu u^\nu + p\Delta^{\mu\nu} + T_{(1)}^{\mu\nu}$, where
$$
  T_{(1)}^{\mu\nu} = ({\cal E} - \epsilon) u^\mu u^\nu + ({\cal P} - p)\Delta^{\mu\nu}
  +({\cal Q}^\mu u^\nu + {\cal Q}^\nu u^\mu) + {\cal T}^{\mu\nu}
$$
is first order in derivatives. The definition (\ref{eq:Smu-def}) then gives 
\begin{align}
  S^\mu_{\rm canon} = s u^\mu - \frac{u_\nu}{T} T_{(1)}^{\mu\nu}\,.
\end{align}
This entropy current is frame-invariant under first-order redefinitions of the hydrodynamic variables. The hydrodynamic equation $u_\nu \partial_\mu T^{\mu\nu} = 0$ implies $\partial_\mu (s u^\mu) = \frac{u_\nu}{T}\partial_\mu T_{(1)}^{\mu\nu}$ (upon using $\partial p = s\partial T$). This immediately gives the divergence of the entropy current in terms of the non-ideal contribution to the energy-momentum tensor,
\begin{align}
  \partial_\mu S^\mu_{\rm canon} = - T_{(1)}^{\mu\nu} \partial_\mu \!\left( \frac{u_\nu}{T} \right)\,.
\end{align}
We want the right-hand side to be non-negative. This requirement will constrain the form of $T_{(1)}^{\mu\nu}$, in other words it will constrain the constitutive relations.
To proceed, note the following. Take any symmetric tensor $X^{\mu\nu}$ and decompose it as in~(\ref{eq:T1}),
\begin{equation}
\label{eq:X1}
  X^{\mu\nu} = {\cal E}_X u^\mu u^\nu + {\cal P}_X \Delta^{\mu\nu} +
     ({\cal Q}_X^\mu u^\nu + {\cal Q}_X^\nu u^\mu) + {\cal T}_X^{\mu\nu}\,.
\end{equation}
where again
\begin{subequations}
\label{eq:EPNqjtX}
\begin{eqnarray}
  && {\cal E}_X \equiv u_\mu u_\nu X^{\mu\nu}\,,\ \ \ \ 
     {\cal P}_X \equiv \frac{1}{d} \Delta_{\mu\nu} X^{\mu\nu}\,,\ \ \ \ 
     {\cal Q}_{X,\mu} \equiv -\Delta_{\mu\alpha} u_\beta X^{\alpha\beta}\,,\\[5pt]
  && {\cal T}_{X,\mu\nu} \equiv \frac12\left(
     \Delta_{\mu\alpha}\Delta_{\nu\beta} + \Delta_{\nu\alpha}\Delta_{\mu\beta}
     -\frac{2}{d} \Delta_{\mu\nu} \Delta_{\alpha\beta}
     \right) X^{\alpha\beta}\,.
\end{eqnarray}
\end{subequations}
The contraction of two symmetric tensors in such a decomposition is
\begin{equation}
\label{eq:XY}
  X_{\mu\nu} Y^{\mu\nu} = {\cal E}_X {\cal E}_Y + d\, {\cal P}_X {\cal P}_Y
  -2 {\cal Q}_{X,\mu} {\cal Q}_Y^\mu + {\cal T}_{X,\mu\nu} {\cal T}_Y^{\mu\nu}\,.
\end{equation}
Now take $X_{\mu\nu} = \frac12 (\partial_\mu(u_\nu/T) + \partial_\nu(u_\mu/T))$, so that $\partial_\mu S^\mu_{\rm canon} = -T_{(1)}^{\mu\nu} X_{\mu\nu}$. For this $X_{\mu\nu}$, the coefficients in the decomposition (\ref{eq:X1}) are readily evaluated to be
$$
  {\cal E}_X = \frac{\dot T}{T^2}\,,\ \ \ \ 
  {\cal P}_X = \frac1d \frac{\partial_\lambda u^\lambda}{T}\,,\ \ \ \ 
  {\cal Q}_X^\mu = -\frac{1}{2T} \left( 
  \frac{\partial^\mu T}{T} + \frac{u^\mu \dot T}{T} + \dot u^\mu \right)\,,\ \ \ \ 
  {\cal T}_X^{\mu\nu} = \frac{1}{2T} \sigma^{\mu\nu}\,.
$$
Using the contraction (\ref{eq:XY}), the divergence of the canonical entropy current is
$$
  T \partial_\mu S^\mu_{\rm canon} = -({\cal E}-\epsilon) \frac{\dot T}{T} - ({\cal P}-p)\partial_\lambda u^\lambda
  - \left( \frac{\partial^\mu T}{T} + \frac{u^\mu \dot T}{T} + \dot u^\mu \right) {\cal Q}_\mu
  - \frac12 \sigma_{\mu\nu} {\cal T}^{\mu\nu}\,.
$$
So far we haven't said anything about the constitutive relations. Now let us substitute the definitions of the transport coefficients in (\ref{eq:EPQTNJ}), which gives
\begin{align}
  T \partial_\mu S^\mu_{\rm canon} 
   = &
  -\varepsilon_1 \!\left( \frac{\dot T}{T} \right)^2
  -\pi_2 \!\left( \partial_\lambda u^\lambda\right)^2
  -\theta_2 \frac{(\Delta^{\mu\alpha}\partial_\alpha T) (\Delta_{\mu\beta}\partial^\beta T)}{T^2}
  -\theta_1 \dot u^\mu \dot u_\mu
  +\frac{\eta}{2} \sigma_{\mu\nu} \sigma^{\mu\nu} \nonumber\\
  & - (\varepsilon_2 + \pi_1) \frac{\dot T}{T} \partial_\lambda u^\lambda
    - (\theta_1 + \theta_2) \frac{\dot u^\mu \Delta_{\mu\nu} \partial^\nu T}{T}\,.
\label{eq:TdSmu}
\end{align}
In the right-hand side of (\ref{eq:TdSmu}), each square in the first line is positive semi-definite.%
\footnote{
This is because $\Delta^{\mu\alpha}\partial_\alpha T$ and $\dot u^\mu$ are transverse to $u_\mu$ and are therefore spacelike. Similarly, $\sigma_{\mu\nu}$ is transverse to $u^\mu$, and so in the coordinates in which $u^\mu=(1,{\bf 0})$ we have $\sigma_{\mu\nu} \sigma^{\mu\nu} = \sigma_{ij} \sigma^{ij}\geqslant 0$.
}
On the other hand, the terms in the second line in (\ref{eq:TdSmu}) can be of either sign. A non-negative entropy production $\partial_\mu S^\mu_{\rm canon} \geqslant 0$ for an arbitrary fluid flow amounts to $\eta\geqslant 0$, together with demanding that the matrices
\begin{align}
  M_s \equiv 
  \begin{pmatrix}
  -\varepsilon_1 & -\frac12(\varepsilon_2 + \pi_1)\\
  -\frac12(\varepsilon_2 + \pi_1) & -\pi_2
  \end{pmatrix} ,\ \ \ \ 
  M_v \equiv 
  \begin{pmatrix}
  -\theta_1 & -\frac12(\theta_1+\theta_2)\\
  -\frac12(\theta_1+\theta_2) & -\theta_2
  \end{pmatrix}
\end{align}
are positive semi-definite. Demanding that the principal minors are non-negative, it is easy to see that $M_v$ is positive semi-definite only if $\theta_1=\theta_2\leqslant0$. Similarly, $M_s$ is positive semi-definite only if $\varepsilon_1\leqslant0$, $\pi_2\leqslant0$, and $4\varepsilon_1 \pi_2 - (\varepsilon_2 + \pi_1)^2\geqslant0$.

One may be tempted to take these constraints on $\varepsilon_{1,2}$, $\pi_{1,2}$, $\theta_{1,2}$ at face value. However, doing so would be incorrect. These constraints follow by demanding that $\partial_\mu S^\mu_{\rm canon} \geqslant 0$ for all solutions to the first-order hydrodynamic equations, both physical (small gradients) and not (large gradients). If the entropy current is to have a physical interpretation, one can only legitimately insist that $\partial_\mu S^\mu_{\rm canon} \geqslant 0$ is true for physical solutions only. Put differently, we have found that a frame-independent entropy current constrains frame-dependent quantities such as $\varepsilon_1$, so something is amiss with the argument. In order to fix the problem, the derivative expansion has to be implemented for the on-shell quantities in the right-hand side of Eq.~(\ref{eq:TdSmu}).

Recall that $\partial_\mu S^\mu_{\rm canon} \geqslant 0$ is only supposed to be true for flows that satisfy first-order hydro equations. The zeroth-order hydrodynamic equations give
\begin{align}
  \frac{\dot T}{T}   = - \vs^2\, \partial_\lambda u^\lambda + O(\partial^2)\,,\ \ \ \ 
  \frac{\Delta_{\mu\lambda} \partial^\lambda T}{T}   = - \dot u_\mu + O(\partial^2)\,,
\end{align}
where $\vs^2 \equiv \partial p/\partial\epsilon = w/(T\partial\epsilon/\partial T)$.
If we now use these in (\ref{eq:TdSmu}), we find
$$
   T \partial_\mu S^\mu_{\rm canon} 
   = 
  \left( -\pi_2 + \vs^2 (\varepsilon_2 {+} \pi_1) - \vs^4 \varepsilon_1 \right) 
  \!\left( \partial_\lambda u^\lambda\right)^2
  +\frac{\eta}{2} \sigma_{\mu\nu} \sigma^{\mu\nu} + O(\partial^3)\,.
$$
The coefficients $\theta_1$ and $\theta_2$ drop out and do not appear in $\partial_\mu S^\mu_{\rm canon}$ at $O(\partial^2)$. If we demand that the right-hand side is positive semi-definite at $O(\partial^2)$ only (assuming that $O(\partial^3)$ terms are much smaller and won't change the sign of the entropy production), we find a more relaxed set of constraints
\begin{equation}
  \eta\geqslant 0\,,\ \ \ \ 
  -\pi_2 + \vs^2 (\varepsilon_2 {+} \pi_1) - \vs^4 \varepsilon_1  \geqslant 0\,.
\label{eq:constraint-2}
\end{equation}
This is consistent with identifying $\zeta = (-\pi_2 + \vs^2 (\varepsilon_2 {+} \pi_1) - \vs^4 \varepsilon_1)$ as the bulk viscosity (which is frame-invariant), as discussed in Sec.~\ref{sec:const-rel}. As expected, the only constraints from the entropy current at one-derivative order are $\eta\geqslant0$ and $\zeta\geqslant0$, i.e.\ positive viscosities give rise to positive entropy production for physical solutions.

\section{Conformal theories}
\label{sec:conformal}
If the microscopic theory happens to be conformally invariant, the conformal symmetry imposes constraints on the constitutive relations in hydrodynamics, see Refs.~\cite{Baier:2007ix, Loganayagam:2008is}. Let us find the constraints on the transport coefficients in Eq.~(\ref{eq:EPQTNJ}) that follow from the conformal symmetry, in $D>2$ spacetime dimensions. In a conformal theory, the energy-momentum tensor is traceless with Weyl weight $D+2$, while the current has Weyl weight $D$. In other words, under the Weyl rescaling of the metric $g_{\mu\nu} = e^{2\phi} \tilde g_{\mu\nu}$ we must have
\begin{align}
\label{eq:TJ-scaling}
  T^{\mu\nu} = e^{-(D+2)\phi}\, \tilde T^{\mu\nu}\,,\ \ \ \ 
  J^\mu = e^{-D \phi} \, \tilde J^\mu\,.
\end{align}
The conformal anomaly can be ignored in one-derivative hydrodynamics. The hydrodynamic variables transform as $u^\mu = e^{-\phi} \tilde u^\mu$, $T = e^{-\phi} \tilde T$, $\mu = e^{-\phi} \tilde\mu$. The three one-derivative scalars $s_1\equiv \dot T/T$, $s_2\equiv \nabla_\lambda u^\lambda$, $s_3 \equiv u^\lambda \partial_\lambda(\mu/T)$ transform as 
\begin{align}
 &  s_1 = e^{-\phi} (\tilde s_1 - \tilde u^\lambda \partial_\lambda \phi)\,,\\
 &  s_2 = e^{-\phi} (\tilde s_2 + (D{-}1) \tilde u^\lambda \partial_\lambda \phi)\,,\\
 &  s_3 = e^{-\phi} \,\tilde s_3\,.
\end{align} 
The one-derivative vectors $v_1^\mu \equiv \dot u^\mu = u^\lambda \nabla_{\lambda} u^\mu$, $v_2^\mu \equiv \frac{\Delta^{\mu\lambda} \partial_\lambda T}{T}$, $v_3^\mu \equiv \Delta^{\mu\lambda} \partial_\lambda(\mu/T)$ transform as 
\begin{align}
  & v_1^\mu \equiv e^{-2\phi} (\tilde v_1^\mu + \tilde\Delta^{\mu\lambda} \partial_\lambda \phi)\,,\\
  & v_2^\mu \equiv e^{-2\phi} (\tilde v_2^\mu - \tilde\Delta^{\mu\lambda} \partial_\lambda \phi)\,,\\
  & v_3^\mu = e^{-2\phi} \tilde v_3^\mu\,.
\end{align}
The one-derivative tensor $\sigma^{\mu\nu}$ transforms as
\begin{align}
  \sigma^{\mu\nu} = e^{-3\phi} \tilde \sigma^{\mu\nu}\,.
\end{align}
Given the constitutive relations (\ref{eq:T1}) and (\ref{eq:EPQTNJ}), the tracelessness of $T^{\mu\nu}$ implies ${\cal E} = (D{-}1){\cal P}$, or
\begin{align}
  \epsilon = (D{-}1)p\,,\ \ \ \ 
  \varepsilon_i = (D{-}1) \pi_i \,.
\end{align}
Thus is a conformal fluid $(\partial p/\partial\epsilon)_n = 1/(D{-}1)$, $(\partial p/\partial n)_\epsilon = 0$. The frame invariants~$f_i$ in Eq.~(\ref{eq:frame-invariants-1}) then all vanish in a conformal fluid, and so does the bulk viscosity in Eq.~(\ref{eq:zetasigma}), $\zeta=0$.
At zero-derivative order, the scaling (\ref{eq:TJ-scaling}) implies 
\begin{align}
  p(T,\mu) = T^D \u{p}(T/\mu)\,,\ \ \ \ 
  n(T,\mu) = T^{D-1} \u{n}(T/\mu)\,,
\end{align}
where $\u{p}$, $\u{n}$ are dimensionless functions of $T/\mu$. At one-derivative order, the scaling (\ref{eq:TJ-scaling}) implies 
\begin{align}
&  \pi_1 = (D{-}1)\pi_2\,,\ \ \ \ 
   \nu_1 = (D{-}1) \nu_2\,,\\
\label{eq:theta-gamma}
&  \theta_1 = \theta_2\,,\ \ \ \ 
  \gamma_1 = \gamma_2\,,
\end{align}
together with 
\begin{align}
  & \pi_i = T^{D-1} \u{\pi_i}(\mu/T)\,,\ \ \ \ 
    \theta_i = T^{D-1} \u{\theta_i}(\mu/T)\,,\ \ \ \ 
    \eta = T^{D-1} \u{\eta}(\mu/T)\,,\\
  & \nu_i = T^{D-2} \u{\nu_i}(\mu/T)\,,\ \ \ \ 
    \gamma_i = T^{D-2} \u{\gamma_i}(\mu/T),
\end{align}
where again $\u{\pi_i}$, $\u{\theta_i}$, $\u{\eta}$, $\u{\nu_i}$, $\u{\gamma_i}$ are dimensionless functions of $\mu/T$. Note that the relations (\ref{eq:theta-gamma}) are consequences of extensivity (see Sec.~\ref{sec:extensivity}), and are true for non-conformal fluids as well.

In particular, in an uncharged conformal fluid the most general one-derivative constitutive relations (\ref{eq:EPQTNJ}) are determined by only four dimensionless numbers $\u{p}$, $\u{\pi_1}$, $\u{\theta_1}$, and $\u{\eta}$, so that
\begin{align}
  T^{\mu\nu} 
  & = T^{D-1}\left( \u{p} T + \u{\pi_1} \frac{\dot T}{T} + \u{\pi_1} \frac{\nabla_\lambda u^\lambda}{D{-}1}  \right) \left( g^{\mu\nu} + D\, u^\mu u^\nu \right)  \nonumber \\
  & + \u{\theta_1} T^{D-1}\left[ \left( \dot u^\mu + \frac{\Delta^{\mu\lambda}\partial_\lambda T}{T}\right)u^\nu + (\mu{\leftrightarrow}\nu)\right] - \u{\eta} T^{D-1} \sigma^{\mu\nu} + O(\partial^2)\,.
\end{align}
The stability and causality of this fluid in $D=4$ dimensions was studied in Ref.~\cite{Bemfica:2017wps}.

\bibliographystyle{JHEP}
\bibliography{hydro-general-biblio}

\end{document}